\documentclass[english,preprint,JIP]{ipsj}

\usepackage{graphicx,xcolor}
\usepackage[fleqn]{amsmath}

\usepackage{array}
\usepackage{xurl}
\usepackage[nobreak]{cite}

\usepackage{threeparttable}
\usepackage{booktabs}

\usepackage[en,cr]{revision}
\IfRevision{\externaldocument{reply_letter/reply_letter}}
\usepackage{revision_def}
\usepackage{revision_ipsj_ieice}

\newcommand{\mymax}{\mathop{\rm max}\limits}

\newcommand{\LifetimeThresholdLong}{L_{t\mbox{\scriptsize -}\mathrm{long}}}
\newcommand{\LifetimeThresholdShort}{L_{t\mbox{\scriptsize -}\mathrm{short}}}

\usepackage{framed}
\usepackage{algorithm}
\usepackage[noend]{algpseudocode}

\newenvironment{ls_algorithmic}
{\begin{algorithmic}[1]\setlength{\baselineskip}{.9\baselineskip}}
{\end{algorithmic}}
\algrenewcommand\algorithmicindent{1.0em}
\algtext*{EndIf}
\algtext*{EndFor}
\algtext*{EndWhile}
\algtext*{EndFunction}
\algnewcommand\algorithmicforeach{\textbf{for each}}
\algdef{S}[FOR]{ForEach}[1]{\algorithmicforeach\ #1\ \algorithmicdo}
\newcommand\CONDITION[2]%
  {\begin{tabular}[t]{@{}l@{}l@{}}
     #1&#2
   \end{tabular}%
  }
\algdef{S}[IF]{IfLongCond}[1]%
  {\algorithmicif\ \CONDITION{#1}{\ \algorithmicthen}}%

\usepackage{siunitx}
\usepackage{myref_en}

\usepackage{color}
\usepackage{colortbl}
\newcommand{\cellem}{\cellcolor[gray]{.8}}

\renewcommand{\thefootnote}{\fnsymbol{footnote}}

\usepackage{latexsym}

\def\Underline{\setbox0\hbox\bgroup\let\\\endUnderline}
\def\endUnderline{\vphantom{y}\egroup\smash{\underline{\box0}}\\}
\def\|{\verb|}

\usepackage[varg]{txfonts}%
\makeatletter%
\input{ot1txtt.fd}
\makeatother%

\begin{document}

\title{Constructing Object Groups Corresponding to Concepts for Recovery of a Summarized Sequence Diagram}

\affiliate{titech}{Tokyo Institute of Technology, Japan}
\affiliate{astem}{Advanced Science, Technology \& Management Research Institute of KYOTO, Japan}

\author{Kunihiro Noda}{titech}[knhr@sa.cs.titech.ac.jp]
\author{Takashi Kobayashi}{titech}[tkobaya@c.titech.ac.jp]
\author{Kiyoshi Agusa}{astem}[agusa@astem.or.jp]

\begin{abstract}
Comprehending the behavior of an object-oriented system solely from its source code is troublesome owing to its dynamism.
To aid comprehension, visualizing program behavior through reverse-engineered sequence diagrams from execution traces is a promising approach.
However, because of the massiveness of traces, recovered diagrams tend to become very large causing scalability issues.

To address these issues, we propose an object grouping technique
 that horizontally summarizes a reverse-engineered sequence diagram.
Our technique constructs object groups based on Pree's meta patterns in which
each group corresponds to a concept in the domain of a subject system.
By visualizing interactions only among important groups, we generate
a summarized sequence diagram depicting a behavioral overview of the system.

Our experiment showed that our technique outperformed the state-of-the-art trace summarization technique in terms of reducing the horizontal size of reverse-engineered sequence diagrams.
Regarding the quality of object grouping, our technique achieved an $F$-score of 0.670 and a Recall of 0.793 on average under the condition of \#lifelines (i.e., the horizontal size of a sequence diagram) $<$ 30, whereas those of the state-of-the-art technique were 0.421 and 0.670, respectively.
The runtime overhead imposed by our technique was \SI{129.2}{\percent} on average, which is relatively smaller than other figures found in the reference literature.

\end{abstract}


\maketitle

\section{Introduction}
\label{sec:introduction}

Software documentation is a vital source of information for program comprehension;
however, such documentation tends to become outdated because numerous modifications can occur after an application is released.
Generating reverse-engineered sequence diagrams from execution traces is a promising approach that can help developers comprehend the behavior of a system.
Because an execution trace contains a vast amount of information, the trace information must be summarized or abstracted in order to generate reverse-engineered sequence diagrams of a reasonable size\tcite{bennett08:_survey_and_evaluat_of_tool,Ghaleb:survey_sd:2018:JSEP}.

The vertical size of a reverse-engineered sequence diagram increases according to the execution time, while the horizontal size increases in proportion to the number of generated objects.
Most existing approaches focus on reducing the vertical size of reverse-engineered sequence diagrams, using strategies such as compacting repetitive behavior\tcite{taniguchi05:_extrac_sequen_diagr_from_execut,myers10:_utiliz_debug_infor_to_compac,jayaraman:2016:SPE:compact_visualization} and dividing an entire trace into several phases\tcite{watanabe08:_featur_level_phase_detec_for,ishio08:_amida,Pirzadeh:2013:SCP:stratified_sampling_phase_detection,Pirzadeh:2011:ICSM:text_mining_phase_detection}.
To improve practicality, horizontal reduction is also important.
Some approaches perform horizontal reduction using techniques such as
grouping objects\tcite{dugerdil10:_autom_gener_of_abstr_views,Toda:2013:APSEC_Companion:grouping_obj_design_pattern} and
visualizing only interactions related to core (important) objects\tcite{noda:2018:IEICETrans:identifying_core}.

In this paper, we propose a technique for reducing the horizontal size of reverse-engineered sequence diagrams, with the goal of helping developers comprehend a behavioral overview of a system.
Our technique constructs object groups based on Pree's meta patterns\tcite{pree94:_meta_patter_means_for_captur,pree95:_desig_patter_for_objec_orien_softw_devel} that are the most primitive design patterns.
In our technique, each group corresponds to a concept in the domain of a subject system.
In conjunction with the existing core object identification technique\tcite{noda:2018:IEICETrans:identifying_core},
we identify important object groups and visualize only intergroup interactions.
Performing these tasks generates a summarized version of a reverse-engineered sequence diagram.
The summarized diagram depicts a behavioral overview of the subject system and is expected to be a valuable tool for developers in an early stage of program comprehension.

To improve maintainability in object-oriented programming,
a concept is often divided into several classes by using design patterns\tcite{gamma95:_desig_patter,pree95:_desig_patter_for_objec_orien_softw_devel}.
For instance, in a game application, the concept of \textit{player} is divided into several state classes such as \textit{normal state} and \textit{death state} by using the GoF (Gang of Four) state pattern.
From the perspective of maintainability, implementing a concept with several classes is beneficial; however, from the perspective of program comprehension, it produces numerous design elements to understand, which can confuse developers and increase the amount of effort required for program comprehension tasks.
Our technique constructs object groups based on Pree's meta patterns, in which each group corresponds to a concept.
By reunifying divided concepts as object groups, we generate a summarized sequence diagram that depicts a behavioral overview of a system at the concept level.

We evaluated the feasibility and effectiveness of our technique with traces generated from various types of open source software.
The results showed that our technique outperformed the state-of-the-art trace summarization technique in terms of horizontal summarization of reverse-engineered sequence diagrams.
With regard to the quality of object grouping,
under the condition of \#lifelines (i.e., the horizontal size of a sequence diagram) being less than 30, the $F$-score and \textit{Recall} of our technique were 0.670 and 0.793 on average, respectively, whereas those of the state-of-the-art technique were 0.421 and 0.670, respectively.
Our technique imposed a runtime overhead of \SI{129.2}{\percent} on average which is relatively small compared with recent scalable dynamic analysis techniques.

\R{a-3-contrib-1}{
The main contributions of this paper are as follows:
\begin{itemize}
  \item We propose an algorithm for constructing object groups based on Pree's meta patterns, where each object group corresponds to a concept.
  This is the first object-grouping algorithm that leverages Pree's meta patterns in the literature, leading to  \revA{many opportunities to summarize object behavior}.
  \item We present a new and effective algorithm for drawing a summarized sequence diagram that depicts a behavioral overview as intergroup interactions among important object groups.
  \item The feasibility and effectiveness of our technique are evaluated using traces generated from various types of open source software.
\end{itemize}
}

This paper is an extended version of our previous work\tcite{Noda:2012:WCRE:meta_patterns}.
The main differences from our previous work are as follows:
\begin{itemize}
  \item Clarification and an extension of the algorithms for meta pattern detection and object grouping.
  \item A quantitative evaluation using traces generated from various open source applications.
\end{itemize}

The remainder of this paper is organized as follows.
In \secref{sec:background}, we briefly provide the background knowledge required for describing our technique.
\secref{sec:proposed_technique} elaborates our trace summarization technique.
We evaluate our technique in \secref{sec:experiment} and discuss the threats to validity in \secref{sec:threats_to_validity}.
\secref{sec:related_work} describes key related work and
\secref{sec:conclusion} concludes this paper.

\section{Background}
\label{sec:background}

\subsection{Pree's Meta Patterns}
\label{sec:background:meta_patterns}

\begin{figure}[tb]
  \centering
  \includegraphics[width=\columnwidth]{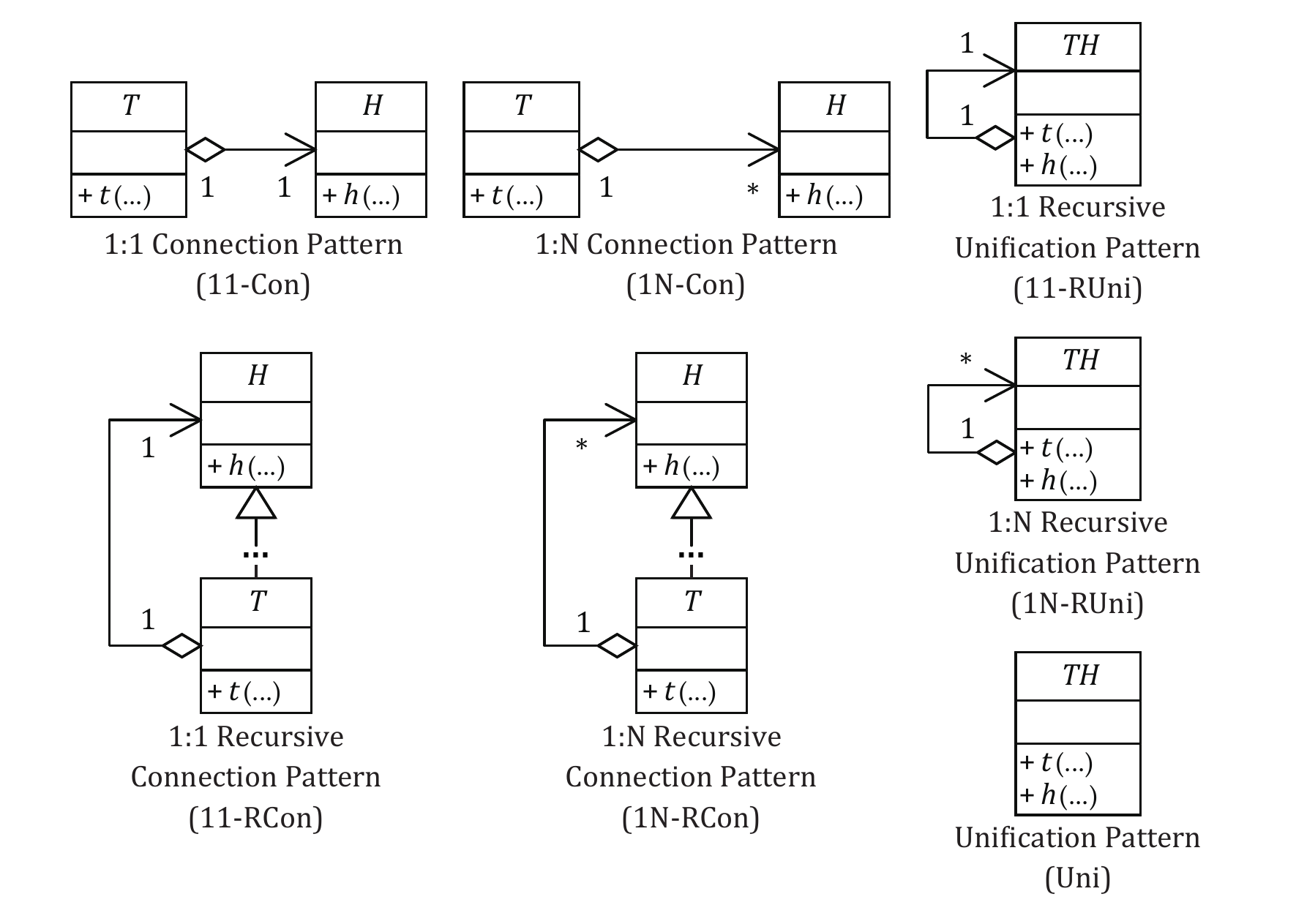}
  \caption{Pree's meta patterns.
   \textit{T, H,} \textit{t}(), and \textit{h}() denote a template class, a hook class, a template method, and a hook method, respectively. \textit{TH} denotes that a template class and a hook class are unified into one class.
   The name in the parentheses below each pattern is the abbreviated name used in this paper.}
  \label{fig:cd_meta_patterns}
\end{figure}

Pree's meta patterns\tcite{pree94:_meta_patter_means_for_captur,pree95:_desig_patter_for_objec_orien_softw_devel} provide
a classification of template-hook structures that are commonly used in object-oriented programming.
\figref{fig:cd_meta_patterns} shows the classification consisting of seven types of template-hook structural patterns.
Pree's meta patterns are the most primitive design patterns.
More concrete design patterns (e.g., the GoF design patterns\tcite{gamma95:_desig_patter}) are realized by using some of the meta patterns.
For example, the GoF decorator pattern is an instance of the 11-RCon pattern.

In our technique, by identifying template-hook objects involved in the same meta pattern,
we construct object groups that each correspond to a concept.

\subsection{Core Object Identification}
\label{sec:background:core_identification}

\R{b-3-background-1}{
We previously presented a core object identification technique (COIT)\tcite{noda:2018:IEICETrans:identifying_core}.
By analyzing reference relations and dynamic properties, the COIT identifies core objects that are important in comprehending the design overview of a system.
The COIT visualizes only interactions related to core objects and thereby generates a summarized sequence diagram that depicts a behavioral overview.
}

The core identification process consists of the following two steps:
 (1) eliminating temporary objects and (2) estimating the importance of each object.

In step (1), the COIT analyzes reference relations and lifetimes for each object to identify temporary objects.
By analyzing reference relations among objects in a manner similar to compilers' escape analysis,
 the COIT identifies the dynamic scope for each object and assigns an escape state to each object.
 The escape states are categorized into three types: GlobalEscape, ReferenceEscape, and Captured.
 GlobalEscape (resp.\ ReferenceEscape) denotes an object that is referenced from another static (resp.\ non-static) object.
 An object is marked as Captured if the object is not referenced from any other objects.

Based on the escape state and lifetime for each object, the COIT identifies an object $o_i$ as a temporary object if it satisfies the following condition.
\begin{align*}
 &(\textit{EscapeState}(o_i) = \textrm{``Captured''}\\
 &\land \textit{Lifetime}(o_i) < \textit{Lifetime}_{\textrm{max}}(\mathcal{O}) \cdot
 \LifetimeThresholdLong)\\
 \lor \ &(\textit{EscapeState}(o_i) = \textrm{``ReferenceEscape''}\\
 &\land \textit{Lifetime}(o_i) < \textit{Lifetime}_{\textrm{max}}(\mathcal{O}) \cdot \LifetimeThresholdShort)
\end{align*}
Here, $\mathcal{O}$ is a set of all the objects.
$\textit{Lifetime}_{\textrm{max}}$ returns the maximum lifetime over all the objects.
$\LifetimeThresholdLong$ and $\LifetimeThresholdShort$ are threshold
factors for deciding whether an object is long-lived, short-lived, or neither.
The above condition means an object $o_i$ is identified as a temporary object if $o_i$ satisfies one of the following conditions: $o_i$ is referenced from no other objects and is not long-lived; $o_i$ is referenced only from other non-static objects and is short-lived.

In step (2), the COIT estimates the importance of each non-temporary object based on the access frequency.
Important objects are expected to be heavily accessed from other objects.
The COIT calculates the importance value of each object as the harmonic mean of the write, read, and method-invocation frequencies.
By building an importance-based object ranking $R = \langle o_1, o_2, ..., o_n \rangle$,
the COIT identifies an object $o_i$ as a core object if the importance of $o_i$ is greater than the threshold $I_t$.

\section{Recovering a Summarized Sequence Diagram by Constructing Object Groups}
\label{sec:proposed_technique}

\R{a-1-1}{
In this section, we describe the details of our technique for recovering a summarized version of a reverse-engineered sequence diagram.
Our technique consists of the following  \revA{components}:
\begin{enumerate}
  \item Meta pattern detection (\secref{sec:proposed_technique:meta_patterns_detection}).
  \item Construction of object groups based on meta patterns (\secref{sec:proposed_technique:object_grouping}).
  \item Visualization of intergroup interactions among important object groups (\secref{sec:proposed_technique:visualization}).
\end{enumerate}
}

\begin{figure}[tb]
  \RFLIPSJ{a-1-2}{}

  \centering
  \includegraphics[width=.7\columnwidth]{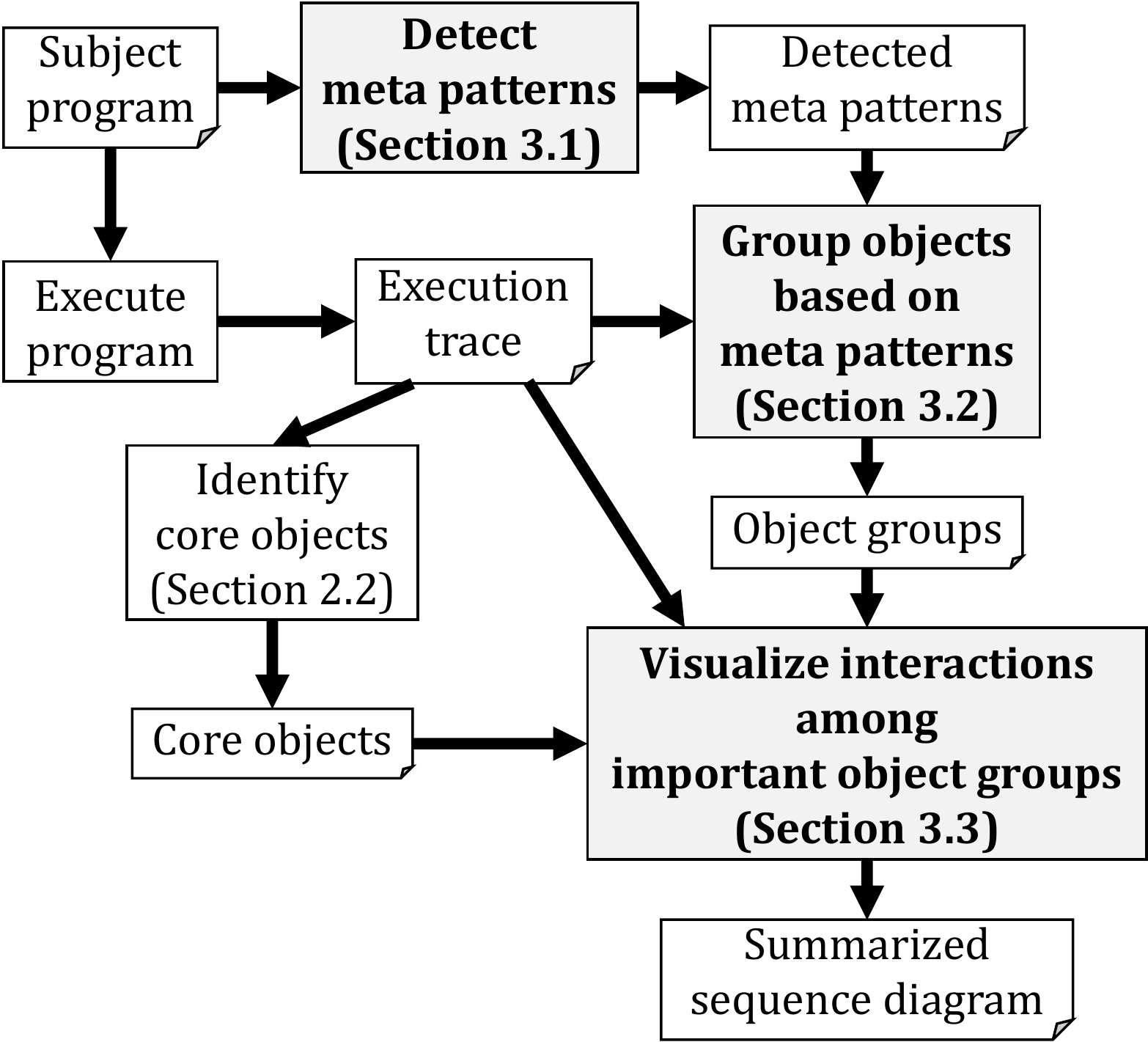}
  \caption{\revA{Overview of our technique.}}
  \label{fig:overview}
\end{figure}

\R{a-1-3}{\revA{
{\figref{fig:overview}} shows an overview of our technique.
Given a subject program, our technique first statically analyzes the source code to detect all meta pattern instances.
By utilizing the detected patterns and an execution trace, it constructs object groups corresponding to concepts in the domain of the subject program.
Then, leveraging a core identification technique, it identifies important object groups.
Finally, intergroup interactions only among important objects are visualized, resulting in a summarized version of a reverse-engineered sequence diagram that depicts a behavioral overview of the subject program.
}}

\subsection{Meta Pattern Detection}
\label{sec:proposed_technique:meta_patterns_detection}

Because Pree's meta patterns employ a structural classification, we can identify the meta patterns via static analysis.
\algoref{alg:meta_patterns_detection} shows our meta pattern detection algorithm.
First, we detect hook methods (l.1).
Then, we identify meta patterns by detecting template methods that invoke the hook methods (l.2).

In hook method detection, 
by visiting all the method declaration nodes in an abstract syntax tree (AST),
we check whether each method overrides/is-overridden-by another method.
If a method overrides/is-overridden-by another method, the method is identified as a hook method (ll.3--15).
Note that we do not treat all constructors and all methods declared in the topmost type (e.g., Object\#equals(...) and Object\#toString(...)) as hook methods.

In meta pattern detection,
by traversing all the method invocation nodes in an AST,
we test whether each method invokes the detected hook methods.
If a method invokes the hook methods, the caller method is treated as a template method, and a meta pattern consisting of the template method and the invoked hook methods is detected (ll.16--26).

\R{a-2-alg1-1}{
We determine the type of each meta pattern by analyzing the relationships
 between the template and hook classes (ll.\revA{27--37}).}
Note that the multiplicity of the relationship between the template and hook classes is determined on the basis of the fields of the template class.
If a template class refers to a hook class via a field whose type is not a subtype of java.util.(Iterable$\mid$Map), 
we determine the multiplicity as 1 (i.e., 1:1 meta patterns);
 otherwise, the multiplicity is treated as N (i.e., 1:N meta patterns).
Although our algorithm for determining multiplicity generates some false positives,
 the final result of object grouping is not affected because
 our grouping algorithm does not require multiplicity information (described in \secref{sec:proposed_technique:object_grouping}).

\begin{algorithm}[tb]
 \caption{meta pattern detection.}\label{alg:meta_patterns_detection}
 \begin{ls_algorithmic}
  \Require a set of source code $\textit{S} = \{s_1, s_2, ..., s_n\}$.
  \Ensure a set of meta-patterns detected $\textit{P} = \{ \textit{p}_1, \textit{p}_2, ..., \textit{p}_m \}$.
  \State $H \gets$ \Call{detectHooks}{$S$}
  \State \Return \Call{detectMetaPatterns}{$S, H$}
  \Statex
  \Function{detectHooks}{$S$}
  \Statex $\triangleright$ \textbf{In:} a set of source code $S$.
  \Statex $\triangleright$ \textbf{Out:} a set of sets $\textit{HS}$ s.t.\ $H \in \textit{HS}$ is a set of hook methods;
  \Statex \hspace{2.8em} $m_h \in H$ overrides or is-overridden-by other methods in $H$.
    \State $\textit{HS} \gets \emptyset$
    \ForEach{$s \in S$}
      \State $a \gets$ the AST of $s$
      \ForEach{method declaration node $n$ in $a$}
        \State $m \gets$ the method declared in $n$
        \If{$(\exists H \in \textit{HS})[m \in H]$}
          \textbf{continue}
        \EndIf
        \State $M_{\textrm{super}} \gets$ all the methods overridden by $m$
        \State $M_{\textrm{sub}} \gets$ all the methods that override $m$
        \If{$M_{\textrm{super}} \neq \emptyset \lor M_{\textrm{sub}} \neq \emptyset $}
          \State $H \gets$ $\{m\} \cup M_{\textrm{super}} \cup M_{\textrm{sub}} $
          \State $\textit{HS} \gets \textit{HS} \cup \{ H \}$
        \EndIf
      \EndFor
    \EndFor
    \State \Return $\textit{HS}$
  \EndFunction
  \Statex
  \Function{detectMetaPatterns}{$S, \textit{HS}$}
  \Statex $\triangleright$ \textbf{In:} a set of source code $S$; a set of sets of hook methods \textit{HS}.
  \Statex $\triangleright$ \textbf{Out:} a set of meta patterns $P$, where $p \in P$ is a triple $\langle m_t, H, t \rangle$,
  \Statex \hspace{2.8em} $m_t$ is a template method, $H$ is a set of hook methods,
  \Statex  \hspace{2.8em} and $t$ is a meta-pattern type.
    \State $P \gets \emptyset$
    \ForEach{$s \in S$}
      \State $a \gets$ the AST of $s$
      \ForEach{method invocation node $n$ in $a$}
        \State $m_i \gets$ the method invoked in $n$
        \State $m_e \gets$ the method enclosing the expression of $n$
        \If{$(\exists H \in \textit{HS})[m_i \in H]$}
          \State $t \gets$ \Call{detectPatternType}{$m_e, m_i$}
          \State $P \gets P \cup \{\langle m_e, H, t \rangle\}$
        \EndIf
      \EndFor
    \EndFor
    \State \Return $P$
  \EndFunction
  \Statex
  \Function{detectPatternType}{$m_t, m_h$}
  \Statex $\triangleright$ \textbf{In:} a template method $m_t$; a hook method $m_h$.
  \Statex $\triangleright$ \textbf{Out:} a meta-pattern type $t = \langle c, m \rangle$, where $c$ is a category name,
  \Statex \hspace{2.8em} and $m$ is multiplicity.
    \State $t_t \gets$ the type declaring $m_t$ (i.e., the template type)
    \State $t_h \gets$ the type declaring $m_h$ (i.e., the hook type)
    \State $\textit{mul} \gets$ the multiplicity of the reference from $t_t$ to $t_h$
    \If{$m_h$ is invoked with the keyword `this.' or `super.'}
      \State \Return $\langle \textrm{``Unification''}, \textit{mul} \rangle$
    \EndIf
    \If{$t_t$ is equal to $t_h$}
      \State \Return $\langle \textrm{``Recursive Unification''}, \textit{mul} \rangle$
    \EndIf
    \If{$t_t$ is one of the subtypes of $t_h$}
      \State \Return $\langle \textrm{``Recursive Connection''}, \textit{mul} \rangle$
    \EndIf
    \State \Return $\langle \textrm{``Connection''}, \textit{mul} \rangle$
  \EndFunction
 \end{ls_algorithmic}
\end{algorithm}

\subsection{Object Grouping based on Meta Patterns}
\label{sec:proposed_technique:object_grouping}

\subsubsection{Overview}
\label{sec:proposed_technique:overview}

We construct object groups by using the meta pattern information detected in 
\secref{sec:proposed_technique:meta_patterns_detection}.

As mentioned in \secref{sec:introduction},
in object-oriented programming, 
a concept is often divided into several classes by using design patterns to improve maintainability.
Thus, we consider a set of template and hook objects involved in the same meta pattern as corresponding to a concept.
Our grouping algorithm identifies template/hook objects for each detected meta pattern and groups those objects.

In a template-hook structure, the template method defines the outline of a process and the hook methods correspond to the details of the process.
Our technique groups the template/hook objects such that the behavior of the template method is retained and the behavior of the hook methods is omitted after grouping.
The resulting summarized sequence diagram thereby depicts a behavioral outline of a system.

We categorize the seven types of Pree's meta patterns into the following three pattern types in terms of behavioral aspects.
\begin{itemize}
  \item Recursive patterns (consisting of 11-RUni, 1N-RUni, 11-RCon, and 1N-RCon).
  \item Connection patterns (consisting of 11-Con and 1N-Con).
  \item Unification pattern (consisting of Uni).
\end{itemize}
For each pattern type, we describe our grouping approach in the following sections.

\subsubsection{Object Grouping for the Recursive Patterns}
\label{sec:proposed_technique:grouping_recursive_pattern}

In a meta pattern of the recursive patterns,
a template object recursively refers to template/hook objects;
 that is, a reference chain consisting of template/hook objects is constructed.
Typically, the template and hook methods have the same method signature.
Once the template method is invoked, template/hook method calls are propagated in the reference chain;
 as a result, a chain of template/hook method calls is constructed.

For example, assume a concept of \textit{file system} that consists of two elements: \textit{directory} and \textit{file}.
Typically, a file system is realized using the GoF composite pattern (i.e., 1N-RCon pattern).
There is a FileBase class that declares some abstract methods for file operations.
A Dir (resp.\ File) class that extends the FileBase class as a template (resp.\ hook) class corresponds to \textit{directory} (resp.\ \textit{file}).

\begin{figure}[tb]
  \centering
  \includegraphics[width=\columnwidth]{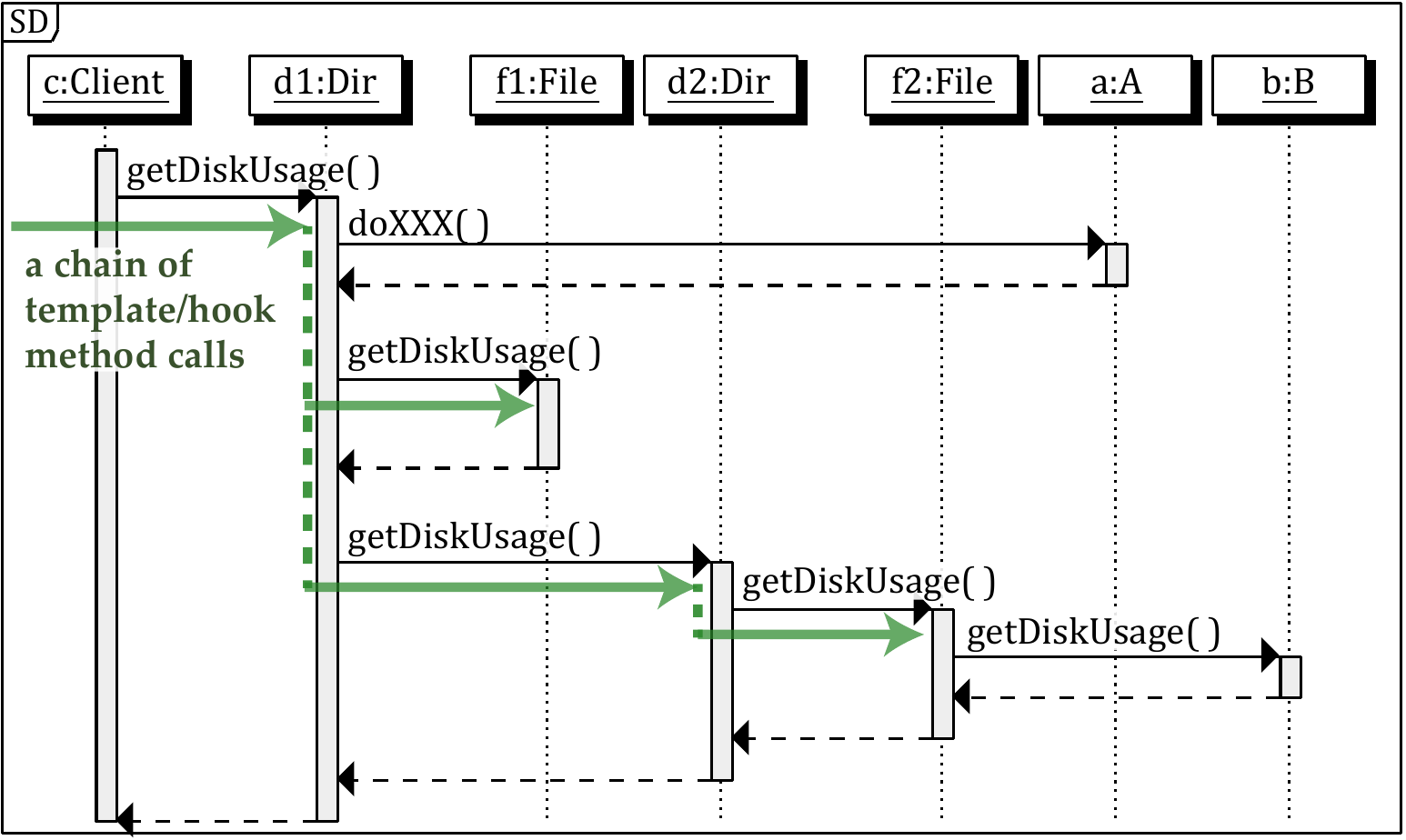}
  \caption{A chain of template and hook method calls. The Dir (resp.\ File) class is a template (resp.\ hook) class. Classes A and B are neither template classes nor hook classes.}
  \label{fig:sd_file_system}
\end{figure}

Assume a method FileBase\#getDiskUsage() that returns disk usage information.
The getDiskUsage() method is overridden in the Directory and File classes.
Once the Dir\#getDiskUsage() (a template method) is invoked, 
 template/hook methods (i.e., Dir\#getDiskUsage and File\#getDiskUsage()) are recursively invoked;
 as a result, a chain of template/hook method calls is constructed as shown in \figref{fig:sd_file_system}.

If the template and hook methods have the same method signature,
 we group all template/hook objects that can be reached by traversing a chain of template/hook method calls.
If the signature of the template method is different from that of the hook methods,
 we group template/hook objects (except for the template object that receives the first message in a chain of template/hook method calls) 
 such that the first template and hook method calls are retained after grouping.
 (Recall that intra-group interactions are omitted in a summarized sequence diagram.)

\begin{figure}[tb]
  \centering
  \includegraphics[width=.6\columnwidth]{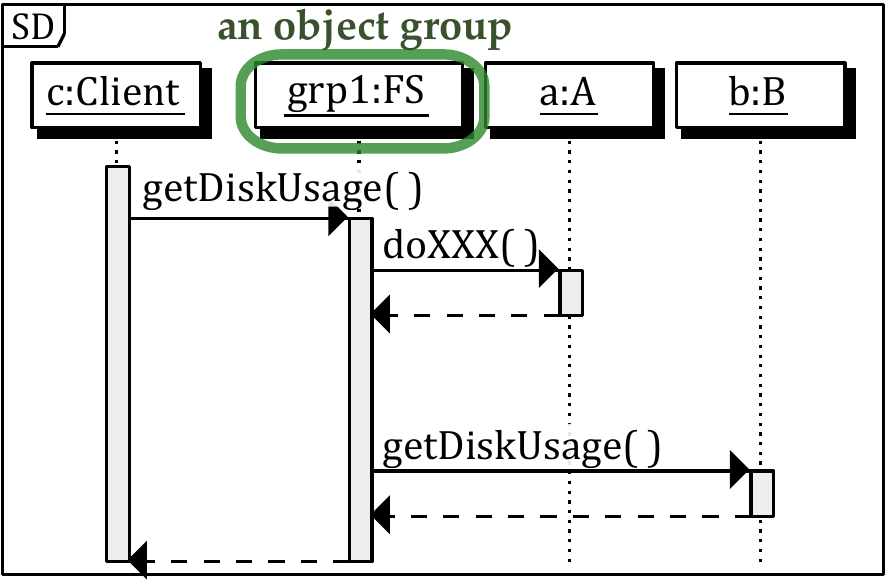}
  \caption{Resulting summarized sequence diagram that depicts an outline of the behavior shown in \figref{fig:sd_file_system}.
   The lifeline named grp1:FS corresponds to the concept of \textit{file system}.
   The lifeline named b:B is not gathered into grp1:FS because class B is neither a template class nor a hook class.}
  \label{fig:sd_file_system_summarized}
\end{figure}

In the case of \textit{file system} shown in \figref{fig:sd_file_system},
all the template/hook objects (i.e., Dir and File instances) are gathered into one group.
\figref{fig:sd_file_system_summarized} shows the resulting summarized sequence diagram that depicts a behavioral outline of \figref{fig:sd_file_system}.
In the summarized diagram, only the first call of getDiskUsage() is retained and successive template/hook method calls are omitted.
Note that although B\#getDiskUsage() has the same signature as the template/hook methods,
 the object of class B is not added into the group because the class B is neither a template class nor a hook class.

\subsubsection{Object Grouping for the Connection Patterns}
\label{sec:proposed_technique:grouping_connection_pattern}

In a meta pattern of the connection patterns, 
a template object refers to hook objects.
Typically, the method signature of the template method is different from those of the hook methods.
Once the template method is invoked, the template object calls the hook method for each hook object.

For example, assume a GUI application built using Model-View-Controller (MVC) architecture.
Once a property of a model object is changed, view objects receive notifications of the property change and update their presentations.
This notification and update mechanisms are typically realized using the GoF observer pattern (i.e., 1N-Con pattern).
A property change in a model is notified by a template method such as Model\#notifyPropertyChanged(...);
 subsequently, for each hook object,
 a hook method such as View\#onPropertyChanged(...) is invoked in the template method, which updates the presentations.

The hook objects involved in a connection pattern,
 whose hook methods are invoked in the template method of the same template object, are gathered into one group.
For example, in a GUI application using MVC architecture mentioned above,
 view objects (i.e., hook objects) that are notified from a model object (i.e., template object) are grouped.
 The resulting summarized sequence diagram depicts interactions between the model object and the group of the view objects.

\subsubsection{Object Grouping for the Unification Pattern}
\label{sec:proposed_technique:grouping_unification_pattern}

In the unification pattern, the template and hook methods belong to the same class.
In other words, a template object also behaves as a hook object.
We do not (cannot) perform any grouping for the unification pattern.
Note that each non-grouped object will be regarded as a group consisting of a single object during visualization (described in \secref{sec:proposed_technique:visualization}).

\subsubsection{Overall Algorithm of Object Grouping}
\label{sec:proposed_technique:overall_algorithm}

We show the overall algorithm of our object grouping in \algoref{alg:object_grouping}.
By \algoref{alg:object_grouping}, object groups are constructed as mentioned in the previous sections.

\begin{algorithm}[tb]
 \RFR{a-2-alg}
 \caption{Object grouping based on meta patterns.}\label{alg:object_grouping}
 \begin{ls_algorithmic}
  \Require an execution trace $\textit{ET} = \langle e_1, e_2, ..., e_l\rangle$;
  \Statex \hspace{1.3em} meta patterns detected $P = \{p_1, p_2, ..., p_m\}$.
  \Ensure a set of object groups $\textit{OGS} = \{ \textit{OG}_1, \textit{OG}_2, ..., \textit{OG}_n \}$.
  \State $T \gets \emptyset$
  \ForEach{\textit{EntryEvent} $e_i \in \textit{ET}$}
    \ForEach{$p \in P$}
      \If{the template method of $p$ is invoked at $e_i$}
        \If{$p$ belongs to \textit{recursive patterns}}
          \State $t \gets$ \Call{detectObjGrpForRecursivePattern}{$e_i, p$}
        \EndIf
        \If{$p$ belongs to \textit{connection patterns}}
          \State $t \gets$ \Call{detectObjGrpForConnectionPattern}{$e_i, p$}
        \EndIf
        \State $T \gets T \cup t$
      \EndIf
    \EndFor
  \EndFor
  \State $\textit{OGS} \gets \emptyset$
  \State $O_{\textrm{all}} \gets$ all the objects generated during the execution
  \ForEach{$o_i \in O_{\textrm{all}}$}
    \ForEach{$p \in P$}
      \State $\textit{OGS} \gets \textit{OGS} \cup \{\{ o \mid o \in \textit{OG} \land \langle o_i, p, \textit{OG} \rangle \in T \}\}$
    \EndFor
  \EndFor
  \State $\textit{OGS} \gets \{\textit{OG} \mid \textit{OG} \in \textit{OGS} \land (\nexists \textit{OG}_i \in \textit{OGS})[\textit{OG} \subset \textit{OG}_i$
  \State \hspace{5.5em} $\land\ \exists o_j \exists o_k \exists p[\langle o_j, p, \textit{OG} \rangle \in T \land \langle o_k, p, \textit{OG}_i \rangle \in T]]\}$
  \State \Return \textit{OGS}
  \Statex
  \Function{detectObjGrpForRecursivePattern}{$e_{\textrm{entry}}$, $p$}
  \Statex $\triangleright$ \textbf{In:} an \textit{EntryEvent} $e_{\textrm{entry}}$; a meta pattern $p =\langle m_t, \textit{H}, t \rangle$.
  \Statex $\triangleright$ \textbf{Out:} a triple $\langle o_t, p, \textit{OG} \rangle$, where $o_t$ is a template object,
  \Statex \hspace{2.8em} $p$ is a meta pattern, and \textit{OG} is an object group.
    \If{$e_{\textrm{entry}}$ is a self-call}
      \State $e_{\textrm{entry}} \gets$ the latest entry event $e_l$
      \Statex \hspace{5.6em} s.t. $e_l$ is not a self-call and occurs before $e_{\textrm{entry}}$
    \EndIf
    \State $\textit{OG} \gets \emptyset$
    \State $m_{\textit{at}} \gets$ the method invoked at $e_{\textrm{entry}}$
    \If{$m_{\textit{at}}$ and $m_h \in \textit{H}$ have the same name}
      \State $\textit{OG} \gets \textit{OG} \cup \{$\Call{calleeObj}{$e_{\textrm{entry}}$}$\}$
    \EndIf
    \State $e_{\textrm{exit}} \gets$ the \textit{ExitEvent} that corresponds to $e_{\textrm{entry}}$
    \State $i_{\textrm{entry}}, i_{\textrm{exit}} \gets$ the indices of $e_{\textrm{entry}}, e_{\textrm{exit}}$ in \textit{ET}, resp.
    \State \revA{$\textit{th}_{\textrm{entry}} \gets$ \Call{threadId}{$e_{\textrm{entry}}$}}
    \State \revA{$\textit{ptrace} \gets$ \Call{trace}{$i_{\textrm{entry}} + 1, i_{\textrm{exit}}, \textit{th}_{\textrm{entry}}$}}
    \State $\textit{cs} \gets$ a new call stack
    \For{\revA{$e_i$ \textbf{in} $\textit{ptrace}$}}
      \If{$e_i$ is an \textit{EntryEvent}}
        \State $\textit{cs}.\textit{push}(e_i)$
        \State $M \gets \{m \mid m$ is the method invoked at $e_k \in \textit{cs}$
        \Statex \hspace{7.9em} $\land\ e_k$ is not a self-call$\}$
        \IfLongCond{$(\forall m \in M)\ [(\exists m_h \in H)$\\ $[m$ has the same name and declaring class as $m_h]]$}
          \State $\textit{OG} \gets \textit{OG} \cup \{$\Call{calleeObj}{$e_i$}$\}$
        \EndIf
      \EndIf
      \If{$e_i$ is an \textit{ExitEvent}}
        \State $\textit{cs}.\textit{pop}(e_i)$
      \EndIf
    \EndFor
    \State \Return $\langle$\Call{calleeObj}{$o_{\textrm{entry}}$} $, p, \textit{OG} \rangle$
  \EndFunction
  \Statex
  \Function{detectObjGrpForConnectionPattern}{$e_{\textrm{entry}}$, $p$}
  \Statex $\triangleright$ The types of the inputs/output are the same
  \Statex \hspace{.6em} as {\sc detectObjGrpForRecursivePattern}.
    \State $e_{\textrm{exit}} \gets$ the \textit{ExitEvent} that corresponds to $e_{\textrm{entry}}$
    \State \revA{$i_{\textrm{entry}}, i_{\textrm{exit}} \gets$ the indices of $e_{\textrm{entry}}, e_{\textrm{exit}}$ in \textit{ET}, resp.}
    \State \revA{$\textit{th}_{\textrm{entry}} \gets$ \Call{threadId}{$e_{\textrm{entry}}$}}
    \State $O_{\textrm{callee}} \gets \{o \mid o$ is an object that receives a message $m$
    \Statex \hspace{0.5em} $\land\ m$ has the same name and declaring class as $m_h \in \textit{H}$
    \Statex \hspace{0.5em} $\land\ m$ comes from \Call{calleeObj}{$e_{\textrm{entry}}$} in \textsc{trace}(\revA{$i_{\textrm{entry}}, i_{\textrm{exit}}, \textit{th}_{\textrm{entry}}$})$\}$
    \State \Return $\langle$\Call{calleeObj}{$o_{\textrm{entry}}$} $, p, O_{\textrm{callee}} \rangle$
  \EndFunction
  \Statex
  \revA{
  \Function{trace}{$i_\textrm{start}, i_\textrm{end}, \textit{threadId}$}
    \State \Return a partial trace $p = \langle e_\textrm{k1}, e_\textrm{k2}, ... \rangle$ extracted from $\textit{ES}$
    \Statex \hspace{4em} s.t.\ $\forall e_{\textrm{k}i} \in \textit{p}\ [i_\textrm{start} \le i \le i_\textrm{end} \land \textsc{threadId}(e_{\textrm{k}i})=\textit{threadId}]$.
  \EndFunction
  }
 \end{ls_algorithmic}
\end{algorithm}

\R{a-2-1-and-b-3-bmodel}{
We assume that an execution trace is represented in the form of an event sequence based on the behavior model (B-model)
 proposed by Noda et al.\tcite{noda:2012:IEICE:reticella}.
The B-model represents the behavior of an object-oriented system.
The B-model consists of event elements such as \textit{EntryEvent} / \textit{ExitEvent} events, 
 which represent ``entry into a constructor/method'' / ``exit from a constructor/method,'' respectively,
 and \textit{VariableDefinition} / \textit{VariableReference} events, 
 which denote that ``a value is assigned to a variable'' / ``a value is read from a variable,'' respectively.
An execution trace can be represented in the form \revA{$\textit{ES}=$} $\langle e_1, e_2, ..., e_n \rangle$,
 where $e_i$ is an event element in the B-model.
 \revA{$\textit{ES}$ is composed of all events from all threads. To support thread-sensitive analyses, each event $e_i$ has as its attribute the id of the thread where the event occurs.}
}

\R{a-2-1-2}{
In \algoref{alg:object_grouping},
once a template method is invoked,
either {\sc detectObjGrpForRecursivePattern} or {\sc detectObjGrpForConnectionPattern} is called according to the pattern type, which constructs an object group (ll.\revA{1--9}).
}

\R{a-2-2}{
For the recursive patterns,
 if the template and hook methods have the same name, the template object is inserted into the resulting group (ll.\revA{23--24}).
The function {\sc calleeObj}($e$) returns the callee object of the specified entry event $e$.
Then, callee objects in a chain of template/hook method calls are added into the resulting group;
 that is, if all the invoked methods in the call stack \textit{cs} are the template/hook methods, all the callee objects are added into the resulting group (ll.\revA{25--37}).
\revA{The function \textsc{threadId}($e$) returns the id of the thread where $e$ occurs.
The function \textsc{trace} returns a partial trace in which each event occurs during the specified period and on the specified thread.}
Note that some code fragments in the template/hook methods might be extracted as private-methods whose names are different from those of the template/hook methods.
To handle such a case, by ignoring self-calls and focusing on inter-object interactions, we detect a chain of template/hook method calls (ll.\revA{19, 20, and 33}).
}

\R{a-2-3}{
For the connection patterns, 
we group hook objects whose hook methods are invoked in the template methods of the same template object (ll.\revA{40--43}).
}

\R{a-2-3-2}{
Our algorithm ignores the types of the return values and the parameters when testing the equality of methods (ll.\revA{23, 34, and 43}) or in other words we equate overloaded methods.
}

\R{a-2-3-3}{
After constructing object groups for each meta pattern,
we unify object groups having the same template object and meta pattern (ll.\revA{12--14});
 e.g., in the GoF state pattern (i.e., 11-Con pattern), 
 several state objects (i.e., the hook objects) that receive messages from the same template object are gathered into one group.
Finally, if an object group is a subset of another group associated with the same meta pattern, we eliminate the subgroup (l.\revA{15}).
}

It is worth noting that the template/hook objects involved in a meta pattern $p$ might also be involved in another meta pattern $p'$.
 Thus, \algoref{alg:object_grouping} constructs soft clusters of objects;
 an object is allowed to belong to multiple object groups.

\R{a-2-4}{}

If a template (or hook) method is invoked via reflection, a chain of template-hook method calls is interrupted by some special method calls (e.g., java.lang.Method\#invoke(...)).
For instance, if a template method is invoked via reflection, we would obtain a method call sequence like ``template(...) $\rightarrow$ template(...) $\rightarrow$ {\it invoke(...) $\rightarrow$ ... (library-method calls) ... $\rightarrow$} template(...) $\rightarrow$ hook(...).''
To deal with this issue, we need to skip such a chain of special library-method calls during object-grouping.
In actual circumstances, library method calls usually are not recorded, meaning that entry/exit-events of library methods do not appear in a B-model event sequence.
This is because (1) developers usually are not interested in detailed interactions within libraries when program comprehension and (2) tracing library behavior incurs a heavy logging overhead, as described in {\secref{sec:weaving_extent}}.
In case developers would like to know detailed interactions within libraries, reflection would be a limitation of our technique.
 However, the main objective of this research is to help developers comprehend behavior specific to their subject system rather than libraries; thus, reflection does not greatly impair the usefulness of our technique.

\subsubsection{Object Grouping for Delegation Patterns}
\label{sec:proposed_technique:grouping_delegation_pattern}

In object-oriented programming,
delegation can be used as an alternative to inheritance;
 as a result, some delegate methods might appear in a chain of template/hook method calls.
 Grouping delegate objects that are the callee of those delegate methods together with template/hook objects might improve the quality of object grouping.

For example, in the case of the \textit{file system} mentioned in \secref{sec:proposed_technique:grouping_recursive_pattern},
 the File class might delegate the disk usage calculation to another class under certain conditions.
In \figref{fig:sd_file_system}, suppose that class B is a delegate of the File class and B\#getDiskUsage() is called as a delegate method in File\#getDiskUsage() under certain conditions.
Because the object of the class B is also an element in the concept of \textit{file system}, the object should be added to the object group named grp1.

Modifying \algoref{alg:object_grouping} as follows, we can group delegate objects together with template/hook objects.
\begin{itemize}
  \item \R{a-2-5}{Relaxing the condition of line \revA{34} as follows:} 
        $(\forall m \in M)\ [(\exists m_h \in H)[m$ has the same name as $m_h]]$
        (i.e., the equality check regarding the declaring class is deleted; if a method is invoked by another method having the same name, we consider the invoked method as a delegate method).
  \item Grouping objects involved in a connection pattern in the same manner used for grouping objects involved in a recursive pattern.
        The only exception is that the template object is never added to the resulting group \R{a-2-6}{(i.e., \revA{ll.23--24} are skipped).}
\end{itemize}

The grouping method that allows (resp.\ disallows) delegate methods to appear in a method chain of template/hook method calls is referred to as ``MP+D'' (resp.\ ``MP'').
We investigate how grouping delegate objects affects performance in \secref{sec:experiment} ($\textrm{RQ}_2$).
Our technique groups delegate objects by default because ``MP+D'' outperforms ``MP'' as described in \secref{sec:experiment:answer_rq2}.

\subsection{Visualizing intergroup Interactions among Important Object Groups}
\label{sec:proposed_technique:visualization}

\R{b-3-coit-referred-1}{
Combined with the core identification technique (COIT)\revB{{\tcite{noda:2018:IEICETrans:identifying_core}}} mentioned in \secref{sec:background:core_identification},
we identify important object groups.
By visualizing intergroup interactions only among the important object groups,
we obtain a summarized version of a reverse-engineered sequence diagram that depicts a behavioral overview of a system.
}

\algoref{alg:drawing_sd} shows our algorithm for drawing a summarized sequence diagram.
Given an importance-based object ranking created by the COIT,
if an object group contains an important object $o_\textrm{imp}$ whose importance is greater than the threshold $I_t$, 
we treat the object group as an important one that should appear as a lifeline in the resulting summarized sequence diagram (ll.1--8).
If no groups contain an important object $o_\textrm{imp}$,
 we create a new object group whose only member is the object $o_\textrm{imp}$ (ll.6--7).

 \begin{algorithm}[tb]
 \caption{Drawing a summarized sequence diagram.}\label{alg:drawing_sd}
 \begin{ls_algorithmic}
  \Require an object ranking $\textit{R} = \langle o_1, o_2, ..., o_m\rangle$,
  \Statex \hspace{1.2em} where the importance of $o_i$ is greater than that of $o_{i+1}$;
  \Statex \hspace{1.2em} a threshold of the importance $I_t$;
  \Statex \hspace{1.2em} all the object groups $\textit{OGS} = \{ \textit{OG}_1, \textit{OG}_2, ..., \textit{OG}_n \}$.
  \State $\textit{OGS}_{\textrm{imp}} \gets \emptyset$
  \For{$i \gets 1$ \textbf{to} $m$}
    \If{\Call{importance}{$o_i$} $< I_t$}
      \State \textbf{break}
    \EndIf
    \State $\textit{S} \gets \{\textit{OG} \mid \textit{OG} \in \textit{OGS} \land o_i \in \textit{OG} \}$
    \If{$\textit{S} = \emptyset$}
      \State $\textit{S} \gets \{\{ o_i \}\}$
    \EndIf
    \State $\textit{OGS}_{\textrm{imp}} \gets \textit{OGS}_{\textrm{imp}} \cup \textit{S}$
  \EndFor
  \State $\textit{TGS} \gets \emptyset$
  \ForEach{object group $\textit{OG} \in \textit{OGS}_{\textrm{imp}}$}
    \State $\textit{TGS} \gets \textit{TGS} \cup \{$\Call{typeNameSet}{\textit{OG}}$\}$
  \EndFor
  \State $\textit{OGS}_\textrm{class} \gets \emptyset$
  \ForEach{$\textit{TG} \in \textit{TGS}$}
    \State $\textit{OG}_\textrm{class} \gets \{o \mid o \in \textit{OG} \land \textit{OG} \in \textit{OGS}_\textrm{imp}$
    \Statex \hspace{7.4em} $\land $ \Call{typeNameSet}{\textit{OG}} $ = \textit{TG}\}$
    \State $\textit{OGS}_\textrm{class} \gets \textit{OGS}_\textrm{class} \cup \{ \textit{OG}_\textrm{class} \}$
  \EndFor
  \State \Call{drawIntergroupInteractions}{$\textit{OGS}_\textrm{class}$}
  \Statex
  \Function{typeNameSet}{\textit{OG}}
    \Statex $\triangleright$ \textbf{In:} an object group \textit{OG}.
    \Statex $\triangleright$ \textbf{Out:} a set of type names.
    \State \Return $\{t \mid t$ is the type name of $o \in G\}$
  \EndFunction
 \end{ls_algorithmic}
\end{algorithm}

There are two types of methods for drawing a sequence diagram: class-level and instance-level.
In a class-level sequence diagram, lifelines having the same types are unified into one lifeline and thereby
the horizontal size of the diagram is reduced.
Meanwhile, an instance-level sequence diagram provides a detailed behavioral view for each object,
 which increases the horizontal size of the diagram.
 A class-level diagram is useful in an early stage of program comprehension.
 As the understanding of a subject system deepens, an instance-level sequence diagram would become more suitable.

In \algoref{alg:drawing_sd}, object groups are converted from instance-level into class-level at lines 9--15.
If an instance-level diagram is needed, one only has to skip lines 9 through 15.

The function {\sc drawIntergroupInteractions} draws intergroup interactions among the given groups.
Each group is visualized as a lifeline in the resulting diagram.
For each object group, a \textit{group-id} and a \textit{group-type-name} are displayed in the box on the top of the lifeline
 where the \textit{group-id} is a unique identifier for the group and the \textit{group-type-name} is the type name of an arbitrary object in the object group.

Because our object grouping constructs soft clusters, an object can belong to multiple groups.
Suppose an object $o$ belongs to multiple groups $G_1, G_2, ..., G_n$.
If an entry event $e$ whose callee object is $o$ occurs, we must determine which of the groups should receive the message corresponding to $e$.
If the message is either the template or hook method of a meta pattern $p$, and $G_i$ is constructed in regard to $p$,
 we send the message to $G_i$;
 otherwise, we send the message to an arbitrary group that contains the object $o$.

$ $ %

A previous study observed that in software maintenance tasks, successful developers (maintainers) first comprehended high-level structures of a system, then prepared a detailed plan of changes{\cite{Robillard:tse:2004:how-effective-investigate-code}}.
Another study stated that the most important concepts of a system are usually implemented by very few key classes, and those key classes would be good starting points to comprehend an unfamiliar system{\cite{zaidman:jsme:2008:SMR:SMR370}}.

Our summarization technique would help developers build initial knowledge on high-level behavioral overviews of an unfamiliar system.
They can in this way obtain a compact view describing what kinds of interactions (events) occur among key concepts, which are important for successful maintenance activities, without investigating tens or hundreds of classes involved in execution scenarios of interest.
Thus, we consider that it is suitable to use our technique at the beginning of maintenance tasks so that maintainers can grasp a big picture of a system and the responsibilities of important concepts, which are required for planning proper changes to be made.

After the initial understanding of high-level overviews is established, developers need to dive deeper into the detailed interactions within each important concept (object group).
At such a stage, they would need other functionalities for facilitating detailed investigation, such as unfolding grouped lifelines or visualizing (extracting) the behavior only of specific classes of interest.
Our technique would not be a valuable aid for those situations.
Also, if a maintenance task requires much finer-grained investigation such as checking the values of specific variables (e.g., debugging faulty behavior caused by incorrect variable values), our technique would not be beneficial to developers.
These situations are outside the scope of our technique and should be supported by different approaches and tools.

\section{Experiment}
\label{sec:experiment}

\subsection{Research Questions and Evaluation Approaches}
\label{sec:rqs}

We address the following research questions through the experiment.

\begin{framed}
 \noindent \textbf{$\textrm{RQ}_1$:}
 How effective is our technique in terms of reducing the horizontal size of reverse-engineered sequence diagrams?
\end{framed}
\noindent \textbf{Motivation:}
We aim to investigate the performance of our technique with respect to reducing the horizontal size of reverse-engineered sequence diagrams, 
which is the primary objective of this work.
If a sequence diagram is small and contains object groups that are important in comprehending a design overview,
the diagram is useful for program comprehension.

\noindent \textbf{Evaluation Approach:}
For each subject system, we extract the ground truth of important concepts from execution scenarios, documents, and tutorials (\secref{sec:experiment:ground_truths}).
Utilizing the ground truth, we evaluate the quality of the resulting object groups by using the evaluation measures described in \secref{sec:experiment:evaluation_measures}.
The better the quality measures are, the more useful the resulting summarized diagram will be for program comprehension.
We evaluate the effectiveness of our technique by investigating the trade-off between the values of the quality measures and the horizontal size (i.e., \#lifelines) of the resulting diagram.

\R{b-1-1-and-b-3-alg-diff-1}{
\noindent 
\revB{
\textbf{Baseline Selection:}
Our primary focus is horizontal summarization of reverse-engineered sequence diagram, while most existing studies focus on vertical summarization.
Possible candidates of baseline techniques in our experiment are the existing techniques{\tcite{munakata09:_ogan,dugerdil10:_autom_gener_of_abstr_views,Toda:2013:APSEC_Companion:grouping_obj_design_pattern,Noda:COMPSAC2017:core_objects,noda:2018:IEICETrans:identifying_core}} that perform horizontal summarization (Details of the techniques are described in {\secref{sec:related_work:horizontal_summarization}}).
}}

\R{b-1-2-and-b-3-alg-diff-1}{\revB{
Because the summarization by OGAN{\tcite{munakata09:_ogan}} is based on user queries that are not required in our technique, we cannot fairly compare OGAN with our technique.
The rest of the candidate techniques{\tcite{dugerdil10:_autom_gener_of_abstr_views,Toda:2013:APSEC_Companion:grouping_obj_design_pattern,Noda:COMPSAC2017:core_objects,noda:2018:IEICETrans:identifying_core}} are fully automated.
Of those, we have to exclude the technique by Dugerdil and Repond{\tcite{dugerdil10:_autom_gener_of_abstr_views}} because their tool implementation is unavailable.
The technique by Toda et al.{\tcite{Toda:2013:APSEC_Companion:grouping_obj_design_pattern}} constructs object groups based on GoF's design pattern usage.
Their work is still at an early stage and incomplete.
For example, grouping algorithms have been defined only for 5 GoF's design patterns; grouping behavior remains undefined for the rest of the 18 patterns.
Also, it is undefined how to group objects that participate in multiple patterns.
Thus, we consider it is better to select as our baseline the studies{\tcite{Noda:COMPSAC2017:core_objects,noda:2018:IEICETrans:identifying_core}} rather than the work by Toda et al.{\tcite{Toda:2013:APSEC_Companion:grouping_obj_design_pattern}}.
Because the technique{\tcite{Noda:COMPSAC2017:core_objects}} was outperformed in the subsequent study{\tcite{noda:2018:IEICETrans:identifying_core}}, we finally selected the technique in the paper{\tcite{noda:2018:IEICETrans:identifying_core}} as our baseline.
}}

\R{b-1-3-and-b-3-alg-diff-1}{
\revB{The baseline technique{\tcite{noda:2018:IEICETrans:identifying_core}} identifies core objects as described in \secref{sec:background:core_identification} and then} constructs object groups based on lifetimes and reference relations.
 \revB{With} an importance-based object ranking $R$ and a threshold of importance $I_t$,
the baseline technique constructs object groups by hard clustering as follows.
For each object $o_\textrm{imp}$ whose importance is greater than $I_t$,
the baseline technique gathers objects $o \in \mathcal{O}$
 where each object $o \in \mathcal{O}$ is (in)directly referenced from $o_\textrm{imp}$ and the lifetime of $o$ is smaller than that of $o_\textrm{imp}$.
Thus, for each important object $o_\textrm{imp}$, the baseline technique constructs an object group such that $o_\textrm{imp}$ and the gathered objects constitute a composition relation.
}

\R{b-3-alg-diff-2}{
\revB{
The core identification algorithm used by our technique proposed in this paper is the same as the baseline.
The difference between our technique and the baseline lies in the algorithms of object-grouping.
Namely our technique constructs important object groups based on Pree's meta patterns usage ({\secref{sec:proposed_technique:object_grouping}}), while the baseline groups core-related objects based on lifetimes and reference relations as described above.
}}

The baseline technique generates an instance-level sequence diagram, whereas our technique generates a class-level sequence diagram.
For comparison, we convert the object groups constructed by the baseline technique into class-level groups by applying ll.9--15 in \algoref{alg:drawing_sd}.

$ $ %

\begin{framed}
 \noindent \textbf{$\textrm{RQ}_2$:}
 How much is performance improved by grouping delegate objects together with template/hook objects?
\end{framed}
\noindent \textbf{Motivation:}
As mentioned in \secref{sec:proposed_technique:grouping_delegation_pattern},
 delegate methods might appear in a chain of template/hook method calls.
In this research question, we investigate how grouping delegate objects together with template/hook objects affects performance.

\noindent \textbf{Evaluation Approach:}
We compare the performance of ``MP+D'' and ``MP'' (see \secref{sec:proposed_technique:grouping_delegation_pattern}).
As $\textrm{RQ}_1$, we investigate the trade-off between the values of the quality measures and the \#lifelines.

$ $ %

\begin{framed}
\RFRIPSJInnerFramed{a-3-rq-renumber-1}%
\noindent \textbf{\revA{$\textrm{RQ}_3$:}}
 How much is the runtime overhead imposed by our technique?
\end{framed}
\noindent \textbf{Motivation:}
Our technique weaves logging codes into a subject system for analyzing runtime information, which causes a runtime overhead.
In terms of practicality, it is highly important to ensure that the runtime overhead will be small.
In this research question, we investigate the overhead imposed by our technique.

\noindent \textbf{Evaluation Approach:}
For each subject system, we measure the execution times of the woven/original programs to calculate the runtime overhead.
We evaluate the overhead through a comparison against the overhead incurred by recent scalable dynamic analysis techniques.

\subsection{Experimental Setup}
\label{sec:experiment:experimental_setup}

\subsubsection{Subject Systems and Execution Scenarios}
\label{sec:experiment:subjects_execution_scenarios}

\R{b-2-1}{\revB{
There are no standard datasets to evaluate trace summarization techniques in the literature;
 thus, it needs to seek subject systems that are suitable for the evaluation of our technique.
The followings are the criteria we set for subject system selection.
\begin{itemize}
  \item Each subject system should have documentation especially on its core aspects, or have well-known concepts, so that we can construct ground truth of core-related object grouping.
  \item The domains of subject systems should be different from one another to improve the external validity of the experiment.
  \item Each subject system should not be a toy example but a realistic application in order to conduct practical evaluation.
  \item Subject systems should be publicly available so that future studies can do experiments with the same dataset and compare their performance with ours.
\end{itemize}
}}

\R{b-2-2}{\revB{
We found the four systems listed in {\tabref{tab:target_systems}} met the above criteria; thus, we select those as our subject systems.
Note that it is practically quite difficult to conduct our experiment on dozens or hundreds of subject systems because we have to manually compose a set of ground truths via careful inspection of each system.
All the four subjects we select are completely different kinds of realistic applications in different domains: a game application (jpacman), an event-driven GUI modeling application (JModeller), a command line application with file I/O (wro4j), and a multi-threaded application involving network communication (JMeter).
Hence, we consider the set of the four systems in {\tabref{tab:target_systems}} as a good subject collection for a practical evaluation of our technique.
}
For each subject, we select a representative execution scenario \revB{as shown in \tabref{tab:target_systems}}.
}

\begin{table*}[tb]
 \centering
 \caption{Subject systems and execution scenarios.}
 \label{tab:target_systems}
 \begin{threeparttable}
  \centering
  \begin{tabular}{ccrp{42em}}
   \hline
   Project& Ver.& KLOC \tnote{5} & Execution scenario \\
   \hline
   jpacman \tnote{1}& SVN r53 & 6.0 & 
	       launch the application;
	       start a new game;
	       move the Pac-Man to the right;
	       have the Pac-Man obtain a power cookie and change its state to the power state;
	       quit the application.\\
   JModeller \tnote{2}& SVN r1015 & 45.5 & 
	       launch the application;
	       add two class figures side by side;
	       add attributes and methods;
         edit the names of the class, attributes, and methods;
         add connectors indicating association, dependency, and inheritance between the classes;
	       quit the application.\\
   wro4j \tnote{3}& 1.7.7 & 34.0 &
	       execute \textit{wro4j-runner} while specifying
	       test resources (*.js and *.css files) as target files,
	       `cssMin' as a pre-processor,
	       and `jsMin' as a post-processor.\\
   JMeter \tnote{4}& 3\_1 & 187.7 &
	       execute the application from the command line (non-GUI mode) with the following settings:
	       sending HTTP requests to a web page of a university;
	       \#threads=5, \#ramp-up=2, and \#loop=2; saving the results into report files.\\
   \hline
  \end{tabular}
  \centering
  \begin{tablenotes}
   \scriptsize
   \item[1] \url{http://code.google.com/p/jpacman/}
   \ \ ${}^{2}$ \url{https://sourceforge.net/p/jhotdraw/svn/HEAD/tree/trunk/}
   \ \ ${}^{3}$ \url{http://wro4j.github.io/wro4j/}\\
   ${}^{4}$ \url{http://jmeter.apache.org/}
   \ \ ${}^{5}$ Test code is excluded.
  \end{tablenotes}
 \end{threeparttable}
\end{table*}

\subsubsection{Ground Truths}
\label{sec:experiment:ground_truths}
For each subject system, we define the ground truth for object grouping.
We extract the important concepts in the domain of each subject system from scenario descriptions and documents.
Then, we examine the source code and identify the types (classes) that implement the extracted concepts.
\tabref{tab:ground_truths} shows the ground truth for each subject system.

jpacman is a Pac-Man game application written in Java.
We define the main entities in the Pac-Man game as the important concepts. These entities include \textit{player} (Pac-Man), \textit{ghost}, and \textit{map}.

JModeller is a modeling application for drawing class diagrams; it is built on top of JHotDraw which is known as a well-designed GUI framework.
We extract the important concepts from an introductory article\footnote{\scriptsize \url{https://www.javaworld.com/article/2074997/swing-gui-programming/become-a-programming-picasso-with-jhotdraw.html}}
that describes the design details of the JModeller application;
 e.g., \textit{class}, which is a primary element in a class diagram, and \textit{connection}, which defines a relationship between two classes.

wro4j is an application used for improving the loading time of a web application.
For example, wro4j provides the functionalities of merging and minifying web resources (e.g., *.js and *.css files).
Documentation\footnote{\scriptsize \url{https://wro4j.readthedocs.io/en/stable/DesignOverview/}} written by developers provides an overview of the design of wro4j.
We consider components appearing in the architecture diagram in the document as the important concepts;
 e.g., \textit{model}, which represents web resources to process, and \textit{pre/post-processors}, which process web resources.

JMeter is an application that provides functionality for measuring the performance of a web application.
The user manual\footnote{\scriptsize \url{https://jmeter.apache.org/usermanual/index.html}} describes the components of JMeter.
We extract major concepts relating to our execution scenario from the user manual.
In addition, an overview of a test-execution flow is provided in the wiki\footnote{\scriptsize \url{https://wiki.apache.org/jmeter/JMeterTestExecution}}.
We also extract important concepts from the description;
 e.g., \textit{test configuration}, which specifies the method for measuring the performance, and \textit{sampler}, which sends requests to target servers.

\begin{table*}[tb]
 \centering
 \caption{Important concepts and types corresponding to the concepts (i.e., ground truths).}
 \label{tab:ground_truths}
 \begin{threeparttable}
  \centering
  \begin{tabular}{ccp{43.5em}}
   \hline
   Project& Concept & Types corresponding to the concept \tnote{1} \\
   \hline
   jpacman & Player & {\it Player, StateManager, Player\$NormalState, Player\$PowerState}\\
           & Ghost  & {\it Ghost, StateManager, Ghost\$NormalState, Ghost\$EatableState, RunningGhostBrain, RedGhostBrain }\\
           & Gem    & {\it Gem, EatGem }\\
           & Map    & {\it Map }\\
           \hline
   JModeller & Class        & {\it ClassFigure, SeparatorFigure, GraphicalCompositeFigure, jmodeller.ClassFigure\$(1$|$4$|$5) [TextFigure],}\\
             &              & {\it RectangleFigure, JModellerClass }\\
             & Layout       & {\it StandardLayouter }\\
             & Connection   & {\it AssociationLineConnection, DependencyLineConnection, InheritanceLineConnection, JModellerClass }\\
             & Tool         & {\it DelegationSelectionTool, TextTool, JModellerApplication\$1 [CreationTool], ConnectionTool, UndoableTool }\\
             & Tool palette & {\it ToolButton }\\
             \hline
   wro4j & Model         & {\it WroModel, Group, Resource }\\
         & WroManager    & {\it WroManager }\\
         & PreProcessor  & {\it PreProcessorExecutor\$2 [DefaultProcessorDecorator], BenchmarkProcessorDecorator,}\\
         &               & {\it ExceptionHandlingProcessorDecorator, SupportAwareProcessorDecorator,}\\
         &               & {\it MinimizeAwareProcessorDecorator, ImportAwareProcessorDecorator, CssMinProcessor, CSSMin }\\
         & PostProcessor & {\it GroupsProcessor\$1 [DefaultProcessorDecorator], BenchmarkProcessorDecorator,}\\
         &               & {\it ExceptionHandlingProcessorDecorator, SupportAwareProcessorDecorator,}\\
         &               & {\it MinimizeAwareProcessorDecorator, ImportAwareProcessorDecorator, JSMinProcessor, JSMin }\\
         & Locator       & {\it StandaloneServletContextUriLocator }\\
         \hline
   JMeter & Test configuration & {\it TestPlan, ThreadGroup, HTTPSamplerProxy, ResultCollector, Arguments, LoopController }\\
          & Sampler            & {\it HTTPSamplerProxy, HTTPHC4Impl }\\
          & Listener           & {\it ResultCollector, Summariser, Summariser\$Totals, SummariserRunningSample }\\
          & Test executer      & {\it StandardJMeterEngine, PreCompiler, JMeterThread }\\
   \hline
  \end{tabular}
  \centering
  \begin{tablenotes}
   \scriptsize
   \item[1] The binary names of the types are shown in the column. The package names of the types are omitted for simplicity. If a type is used as an anonymous class in the code, the name of the base class to extend is shown in the following brackets.
  \end{tablenotes}
 \end{threeparttable}
\end{table*}

\subsubsection{Evaluation Measures}
\label{sec:experiment:evaluation_measures}

One commonly used external quality measure for evaluating cluster quality is the F-measure.
We also use the F-measure to evaluate the quality of object grouping.
We calculate the F-measure in the same manner as Steinbach et al.\tcite{Steinbach:KDDWorkshop:2000:comparison_clustering_tech,Steinbach:TechReport:2000:comparison_clustering_tech}.

Let $\textit{RS} = \{R_1, R_2, ..., R_n \}$ be the ground truth (the reference set) of grouping, where $R_i (1 \le i \le n)$ is a set of type names defined in \tabref{tab:ground_truths}.
Let $\textit{OGS}_\textrm{class} = \{\textit{OG}_1, \textit{OG}_2, ..., \textit{OG}_m \}$ be the resulting groups of our technique or the baseline technique,
 where $\textit{OG}_j (1 \le j \le m)$ is a set of objects ($\textit{OGS}_\textrm{class}$ is obtained at l.15 in \algoref{alg:drawing_sd}).
 Let $\textit{TS} = \{ T_1, T_2, ..., T_m \}$ be the resulting groups of type names, where $T_i = \{t \mid t$ is the type name of $o \in \textit{OG}_j \land \textit{OG}_j \in \textit{OGS}_\textrm{class} \}$.

First, for the reference group $\textit{R}_i$ and the resulting group $\textit{T}_j$, the \textit{Recall} and \textit{Precision} are defined as follows:
\begin{align*}
\textit{Recall}(i, j) &= | \textit{R}_i \cap \textit{T}_j |\ /\ |\textit{R}_i|\,, \\
\textit{Precision}(i, j) &= | \textit{R}_i \cap \textit{T}_j |\ /\ |\textit{T}_j|.
\end{align*}
Then, the $F$ score for the reference group $\textit{R}_i$ and the resulting group $\textit{T}_j$ is defined as follows:
\begin{align*}
\textit{F}(i, j) = \frac{2 \cdot \textit{Recall}(i, j) \cdot \textit{Precision}(i, j)}{\textit{Recall}(i, j) + \textit{Precision}(i, j)}.
\end{align*}
Finally, the $F$ score for the reference set $\textit{RS}$ and the grouping result \textit{TS} is calculated as follows:
\begin{align*}
n &= \sum_{R_i \in\, \textit{RS}} |R_i|\,,\\
\textit{F} &= \sum_{R_i \in\, \textit{RS}} \frac{|R_i|}{n} \mymax_{T_j \in\, \textit{TS}} F(i, j).
\end{align*}
The $\textrm{max}_{T_j \in\, \textit{TS}} F(i, j)$ is the maximum value of $F(i, j)$ over all the elements in \textit{TS} and $R_i$.
The $F$ is the weighted average of $\textrm{max}_{T_j \in\, \textit{TS}} F(i, j)$ over all the elements in $\textit{RS}$.

We investigate the trade-off between the $F$ score and the horizontal size of the resulting sequence diagram (i.e., \#lifelines).
Note that, by definition, the $F$ score easily reaches 1 if we create the power set of the set of all the objects.
Therefore, using only the $F$ score is inappropriate for evaluating performance,
 and we must investigate the trade-off between the $F$ score and \#lifelines.

The more the objects of the important types shown in \tabref{tab:ground_truths} are contained in the resulting sequence diagram,
the more useful the resulting diagram will be for program comprehension.
Thus, we also evaluate the rate of the important types contained in the resulting diagram (i.e., the recall).
The \textit{Recall} for the reference set $\textit{RS}$ and the grouping result \textit{TS} is calculated in the same manner as $F$:
\begin{align*}
\textit{Recall} &= \sum_{R_i \in\, \textit{RS}} \frac{|R_i|}{n} \mymax_{T_j \in\, \textit{TS}} \textit{Recall}(i, j)\,.
\end{align*}

\subsubsection{Weaving Extent}
\label{sec:weaving_extent}

\R{a-2-weaving-1}{
Our technique requires results of meta pattern detection (\secref{sec:proposed_technique:meta_patterns_detection})
 and an execution trace.
 The execution trace must contain information regarding method-entries/exits, objects, and field accesses.
 Note that the field access information is used for obtaining an importance-based object ranking $R$ (see \algoref{alg:drawing_sd}).
 We use \textit{SELogger}, which is a part of REMViewer\tcite{Matsumura:2014:ICPC:remviewer_selogger}, for recording execution traces.
}

\R{a-2-and-b-3-weaving}{
We aim to help developers comprehend the behavior specific to the domain of a subject system.
For this reason, we weave logging codes only into the classes defined in the subject system. In other words, libraries are not instrumented.
As the only exception, collection libraries are instrumented in order to analyze reference relations more precisely
 (details are provided in the paper describing the baseline technique\revB{{\tcite{noda:2018:IEICETrans:identifying_core}}}).
This weaving condition is realistic in terms of avoiding a heavy logging overhead.
}

\subsection{Results}
\label{sec:results}

The recorded runtime information is shown in \tabref{tab:runtime_information}.
The numbers of messages and objects affect the size of a reverse-engineered sequence diagram.
The number of lifelines (i.e., the horizontal size) of a class-level (resp.\ instance-level) reverse-engineered sequence diagram is equal to the number of loaded classes (resp.\ objects).

\begin{table}[tb]
 \centering
 \caption{Recorded runtime information.}
 \label{tab:runtime_information}
  \begin{tabular}{crrrr}
   \hline
   Project& Total events & Messages & Loaded classes & Objects \\
   \hline
   jpacman & 19,024,287 & 8,387,304 & 57 & 1,273 \\
   JModeller & 1,074,279 & 568,940 & 153 & 6,993 \\
   wro4j & 311,192 & 147,858 & 295 & 1,504 \\
   JMeter & 630,969 & 333,190 & 372 & 5,562 \\
   \hline
  \end{tabular}
\end{table}

\subsubsection{Answer to $\textrm{RQ}_1$}

\begin{figure}[tb]
  \centering
  \includegraphics[width=.85\columnwidth]{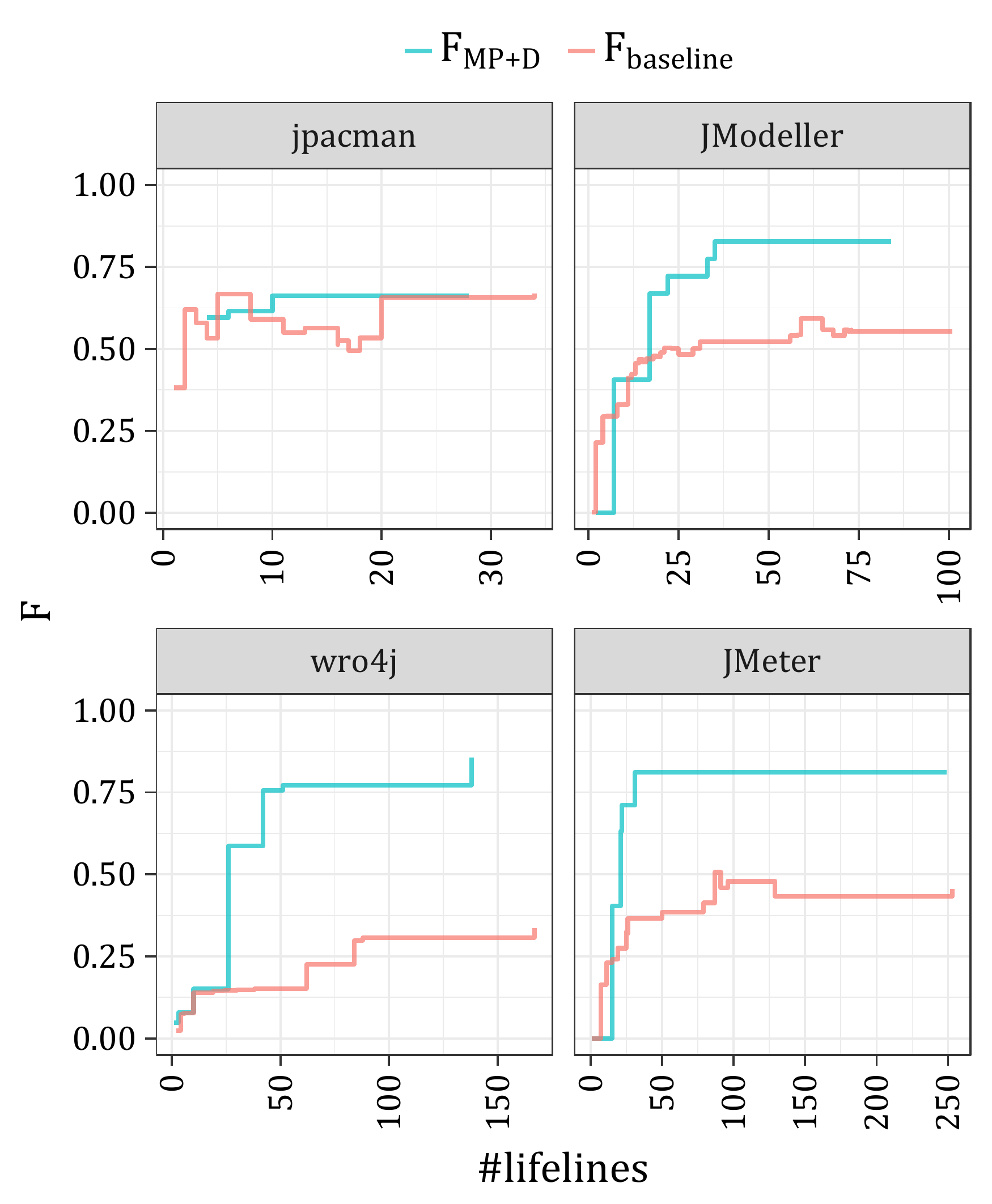}
  \caption{Performance of our technique and the baseline technique ($F$ score vs.\ \#lifelines).}
  \label{fig:performance_ours_baseline_f1}
\end{figure}

\begin{figure}[tb]
  \centering
  \includegraphics[width=.85\columnwidth]{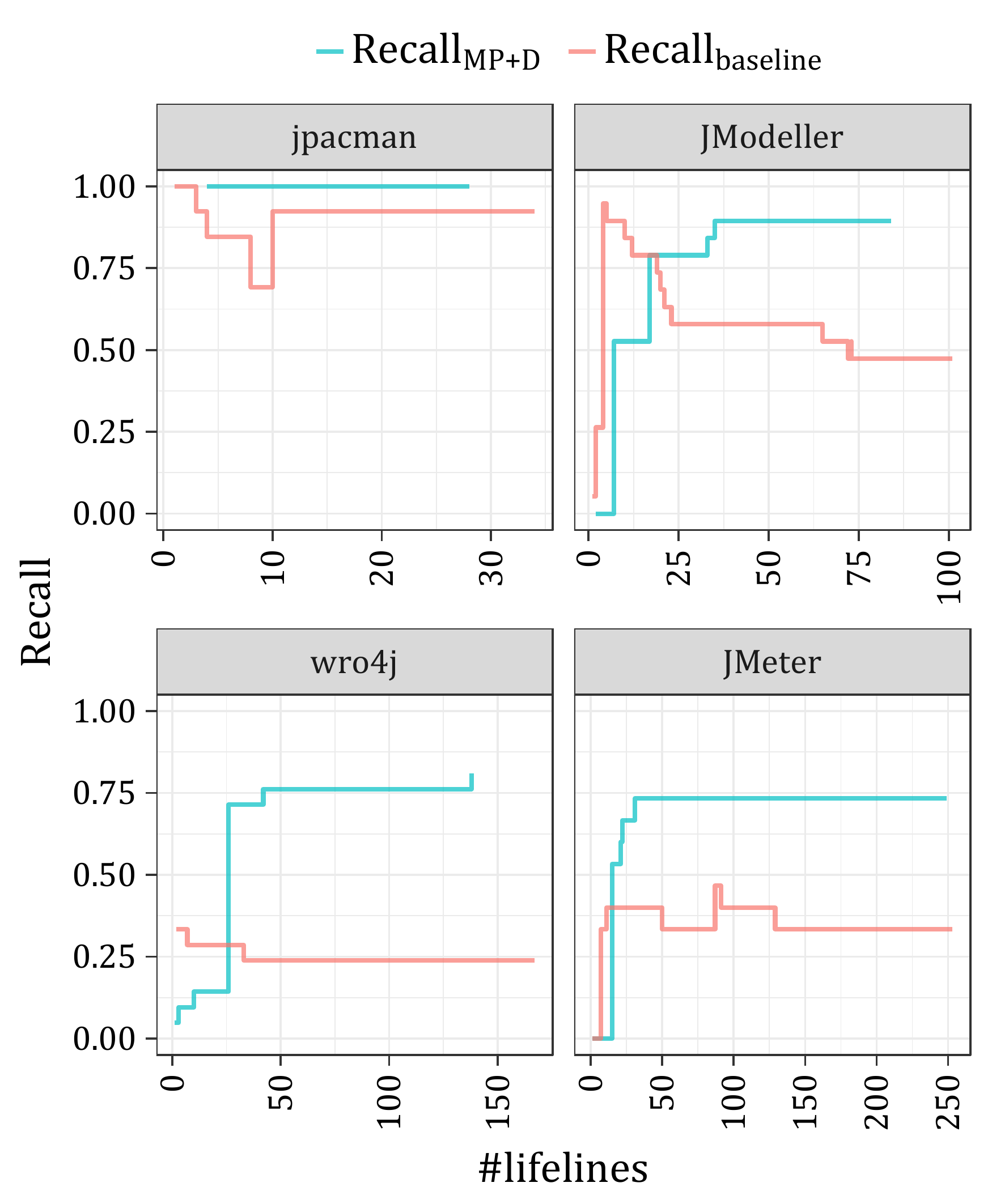}
  \caption{Performance of our technique and the baseline technique ($\textit{Recall}$ vs.\ \#lifelines).}
  \label{fig:performance_ours_baseline_max_recall}
\end{figure}

Varying the threshold $I_t$ (an input of \algoref{alg:drawing_sd})
 causes the resulting object groups to change, which affects the values of the quality measures (i.e., the $F$ score and \textit{Recall}) and \#lifelines.
\figref{fig:performance_ours_baseline_f1} and \figref{fig:performance_ours_baseline_max_recall} show
the trade-off relationship between the quality measures and \#lifelines.

In our technique, the entire set of resulting object groups (i.e., the output \textit{OGS} of \algoref{alg:object_grouping}) is independent of the value of threshold $I_t$.
As the value of $I_t$ decreases, the number of groups displayed in the resulting diagram increases.
Thus, as \#lifelines is increased, the values of the quality measures monotonically increase.

On the other hand, in the baseline technique, the entire set of resulting object groups depends on the value of $I_t$.
After the value of $I_t$ is determined, the baseline technique constructs a set of object groups from scratch.
Thus, the values of the quality measures do not monotonically increase along with the \#lifelines.
For instance, if an object $o$, which (in)directly refers to numerous objects (i.e., $o$ is close to the root object in reference relations), ranks as one of the most important objects, many objects are gathered into the same group, which causes the \textit{Recall} value to increase.
Because the baseline technique constructs hard clusters of objects, reducing the value of $I_t$ increases the number of groups (i.e., causes the large object group containing $o$ to be decomposed);
 this might cause the \textit{Recall} value to decrease, which also affects the $F$ score (e.g., the JModeller case).

As shown in \figref{fig:performance_ours_baseline_f1} and \figref{fig:performance_ours_baseline_max_recall},
 in most of the subject systems, the values of our technique's quality measures increase significantly at the portion of the graph with the fewest \#lifelines,
 and our technique outperforms the baseline technique at \#lifelines $> 25$.
 
A sequence diagram with many \#lifelines \R{a-4-1}{ \revA{(e.g., $>$ 100)}} is not suitable for practical use because it requires significant effort for developers to comprehend the content of the diagram.
Thus, we focus on the area where \#lifelines is less than 30 (i.e., small enough for manual investigation) in \figref{fig:performance_ours_baseline_f1} and \figref{fig:performance_ours_baseline_max_recall}.
The maximum performance under the condition of \#lifelines $< 30$ is shown as \tabref{tab:performance_class_level}.
\tabref{tab:performance_class_level} shows that our technique (the ``MP+D'' column) outperforms the baseline technique for almost all cases.
Our technique achieved an $F$ score (resp.\ a \textit{Recall}) of 0.670 (resp.\ 0.793) on average, which is 0.249 (resp.\ 0.123) higher than that of the baseline technique.
Our technique is therefore more effective compared with the baseline in terms of reducing the horizontal size of reverse-engineered sequence diagrams.

\begin{table}[tb]
 \RFLIPSJTable{a-4-2}{}
 \centering
 \caption{Maximum performance under the condition of  \revA{\#lifelines $<$ 30}. Each shaded cell indicates the maximum performance value for each project.}
 \label{tab:performance_class_level}
  \begin{tabular}{ccccccc}
   \hline
   & \multicolumn{2}{c}{MP+D} & \multicolumn{2}{c}{MP} & \multicolumn{2}{c}{baseline}\\
   \addlinespace[-2pt] \cmidrule(r{0.4em}){2-3} \cmidrule(r{0.3em}){4-5} \cmidrule(l{0.2em}){6-7} \addlinespace[-1.5pt] 
   Project& $F$ & $\textit{Recall}$ & $F$ & $\textit{Recall}$ & $F$ & $\textit{Recall}$ \\
   \hline
   jpacman    & 0.662 &\cellem 1.000 & 0.636 & 0.538 &\cellem 0.667 &\cellem 1.000 \\
   JModeller  &\cellem 0.722 & 0.789 & 0.655 & 0.632 & 0.503 &\cellem 0.947 \\
   wro4j      &\cellem 0.586 &\cellem 0.714 &\cellem 0.586 &\cellem 0.714 & 0.146 & 0.333 \\
   JMeter     & 0.711 &\cellem 0.667 &\cellem 0.767 &\cellem 0.667 & 0.366 & 0.400 \\
   \hline
   Average    &\cellem 0.670 &\cellem 0.793 & 0.661 & 0.638 & 0.421 & 0.670 \\
   \hline
  \end{tabular}
\end{table}

\begin{framed}
 \noindent
 Our technique achieved an $F$ score (resp.\ a \textit{Recall}) of 0.670 (resp.\ 0.793) on average, under the condition of \#lifelines $< 30$ (which is acceptable for manual investigation); this is 0.249 (resp.\ 0.123) higher than that of the baseline technique.
 Thus, our technique is more effective than the baseline technique in terms of reducing horizontal size of reverse-engineered sequence diagrams.
\end{framed}

\subsubsection{Answer to $\textrm{RQ}_2$}
\label{sec:experiment:answer_rq2}

\begin{figure}[tb]
  \centering
  \includegraphics[width=.85\columnwidth]{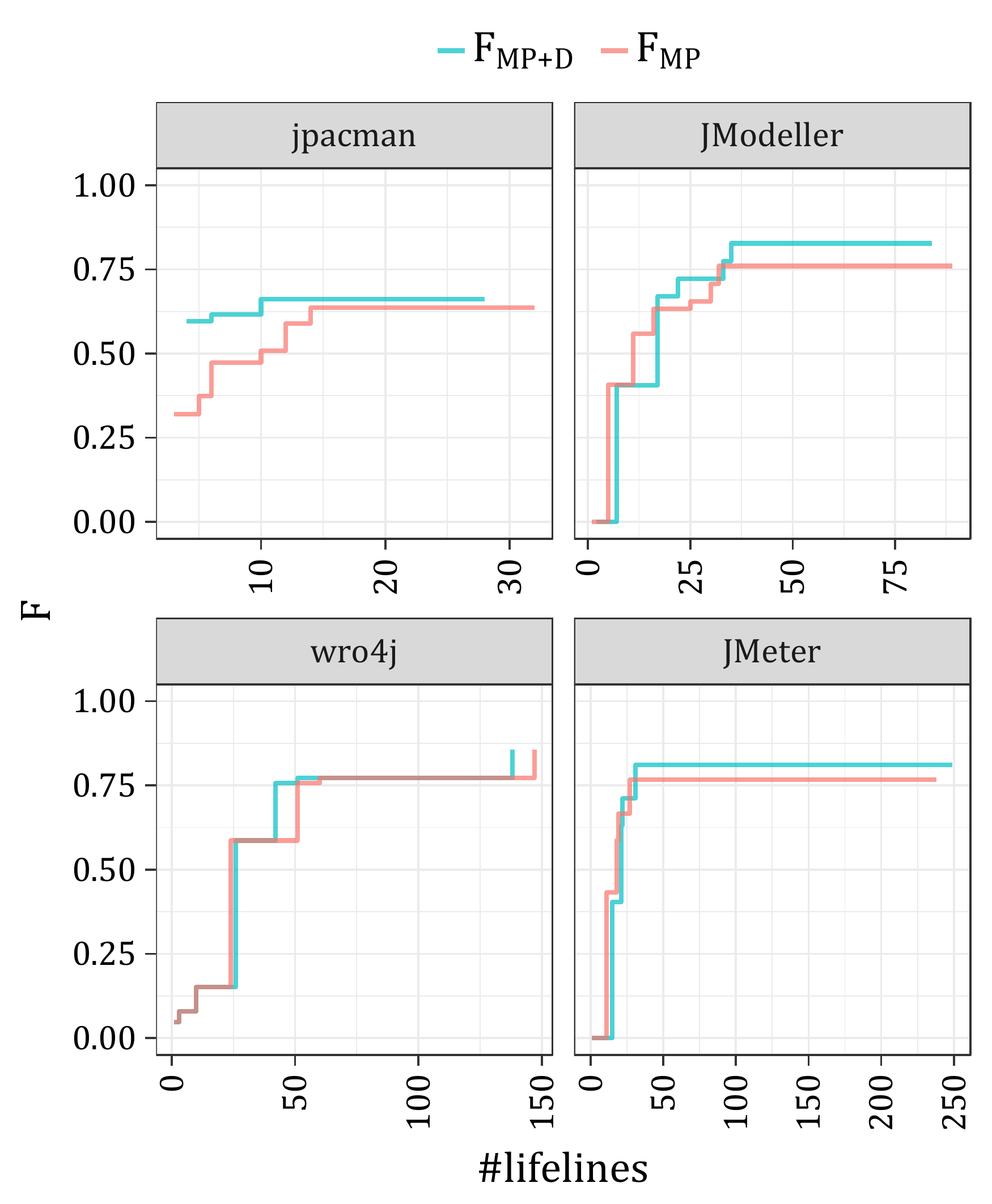}
  \caption{Effect of allowing delegate methods in a chain of template/hook method calls ($F$ score vs.\ \#lifelines).}
  \label{fig:performance_delegation_on_off_f}
\end{figure}

\begin{figure}[tb]
  \centering
  \includegraphics[width=.85\columnwidth]{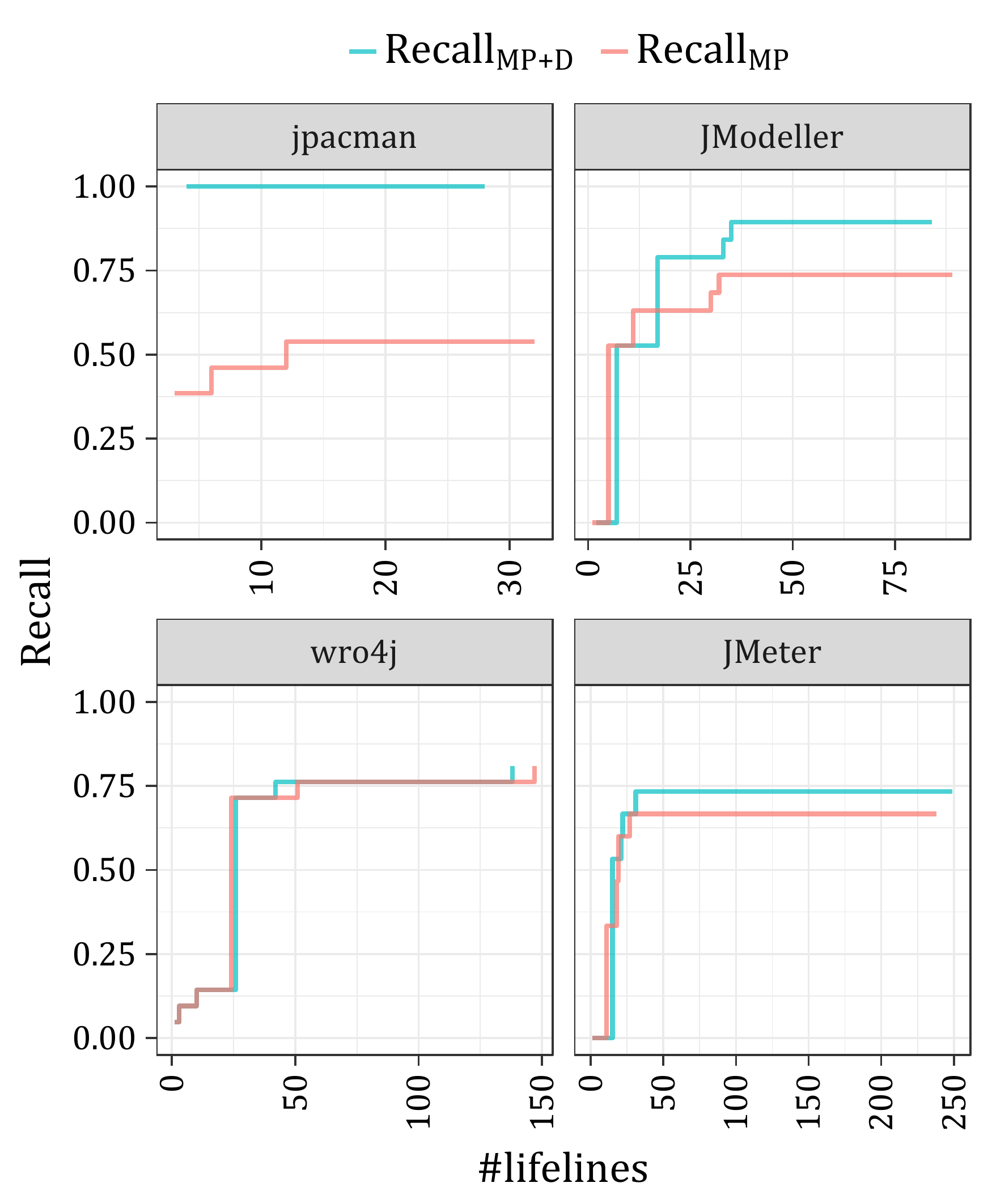}
  \caption{Effect of allowing delegate methods in a chain of template/hook method calls (\textit{Recall} vs.\ \#lifelines).}
  \label{fig:performance_delegation_on_off_max_recall}
\end{figure}

We show the performance of ``MP+D'' and ``MP'' (see \secref{sec:proposed_technique:grouping_delegation_pattern})
 in \figref{fig:performance_delegation_on_off_f} and \figref{fig:performance_delegation_on_off_max_recall}.
\figref{fig:performance_delegation_on_off_f} (resp.\ \figref{fig:performance_delegation_on_off_max_recall}) shows the trade-off relationship between the $F$ score (resp.\ the \textit{Recall}) and \#lifelines.
As shown in \figref{fig:performance_delegation_on_off_f} and \figref{fig:performance_delegation_on_off_max_recall},
 allowing delegate methods in a chain of template/hook method calls (i.e., ``MP+D'') tends to improve the performance of our technique.

\tabref{tab:performance_class_level} shows the maximum performance when there are fewer than 30 \#lifelines, which is a size acceptable for manual investigation.
As shown in \tabref{tab:performance_class_level}, the $F$ score (resp.\ the \textit{Recall}) of ``MP+D'' is 0.009 (resp.\ 0.155) higher than that of ``MP'' on average.
For all the subject systems, allowing delegate methods in a chain of template/hook method calls increased the number of opportunities for object grouping.
Moreover, grouping delegate objects together with template/hook objects improved performance in all cases (except for the $F$ score in the JMeter case);
 i.e., the grouping of delegate objects rarely impaired the performance.

\begin{framed}
 \noindent
 Grouping delegate objects together with template/hook objects improved performance in almost all cases.
 Under the condition of \#lifelines $< 30$, the $F$ score (resp.\ the \textit{Recall}) of ``MP+D'' is 0.009 (resp.\ 0.155) higher than that of ``MP'' on average.
\end{framed}

\R[hideref]{a-3-answer-2}{\revA{
Regarding $\textrm{RQ}_1$ and $\textrm{RQ}_2$, the prevalence of meta patterns and template-hook objects deserves detailed investigation.
Our technique constructs object groups based on meta patterns; therefore,
the number of opportunities for object grouping depends on the number of template/hook methods and objects in subject systems.
If meta patterns are seldom used in subject systems, the number of object groups constructed could be small;
 this might lessen the effectiveness of our technique.
In the following, we show the results of a detailed investigation into the prevalence of meta patterns and template-hook objects.
}}

\R[hideref]{a-3-answer-3}{
\tabref{tab:patterns_rate_in_code} and \tabref{tab:design_patterns_count} show
 the number of meta patterns and delegations for each subject system.
\tabref{tab:patterns_rate_in_code} shows the number of method invocations (i.e., \#pairs of caller and callee methods) that are statically detected (e.g., in the jpacman case, there are 62 pairs of template and hook methods).
\tabref{tab:design_patterns_count} shows the number of pairs of template and hook methods for each pattern type.
}

As shown in \tabref{tab:patterns_rate_in_code},
 \SI{33.2}{\percent} of all method invocations are hook/delegate method invocations on average.
There are a large number of meta patterns and delegations in the subject systems.
Moreover, as shown in \tabref{tab:design_patterns_count}, the unification pattern and connection patterns tend to be used more frequently than the other patterns.

\begin{table}[tb]
 \centering
 \caption{Numbers of method invocations, meta-patterns, and delegations.
 The ``All'' column shows the total number of method invocations in each subject system.
 The ``MP'' column shows the number of meta-patterns (i.e., \#pairs of template and invoked hook methods).
 The ``Delegation'' column shows the number of delegations (i.e., \#pairs of delegating and delegated methods) in which the delegate method is not a hook method.
 Each number is statically counted. Static methods, methods declared in java.lang.Object, and library methods are excluded.
 The ``Rate of MP+D'' is calculated by $(\textrm{``MP''} + \textrm{``Delegation''}) / \textrm{``All''} \cdot 100$.
 }
 \label{tab:patterns_rate_in_code}
  \begin{tabular}{crrrr}
   \hline
   Project& All & MP & Delegation & Rate of MP+D [\%] \\
   \hline
   jpacman    & 425     & 62    & 8   & 16.5 \\
   JModeller  & 7,627   & 3,649 & 100 & 49.2 \\
   wro4j      & 5,035   & 1,837 & 97  & 38.4 \\
   JMeter     & 24,477  & 6,560 & 480 & 28.8 \\
   \hline
   Average    & 9,391   & 3,027 & 171 & 33.2 \\
   \hline
  \end{tabular}
\end{table}

\begin{table}[tb]
 \centering
 \caption{Number of meta-patterns detected.}
 \label{tab:design_patterns_count}
  \begin{tabular}{crrrrrrrr}
   \hline
   \addlinespace[2pt]
   Project& \rotatebox{90}{Uni} & \rotatebox{90}{11--RUni} & \rotatebox{90}{1N--RUni}
             & \rotatebox{90}{11--Con} & \rotatebox{90}{1N--Con} & \rotatebox{90}{11--RCon} & \rotatebox{90}{1N--RCon}\\
   \hline
   jpacman    & 10 & 0 & 0 & 4 & 42 & 0 & 6 \\
   JModeller  & 1,596 & 0 & 9 & 336 & 1,393 & 143 & 172 \\
   wro4j      & 309 & 0 & 1 & 727 & 664 & 48 & 88 \\
   JMeter     & 2,214 & 3 & 13 & 687 & 2,887 & 335 & 421 \\
   \hline
  \end{tabular}
\end{table}

We show the numbers of grouped and non-grouped objects in \tabref{tab:grouped_objects_count}.
In \tabref{tab:grouped_objects_count}, the number outside (resp.\ inside) the parentheses shows the number of non-temporary objects (resp.\ all the objects).
During a program execution, numerous temporary objects are generated and those are not important for program comprehension.
We identified temporary objects using the baseline technique\tcite{noda:2018:IEICETrans:identifying_core}.
Note that those temporary objects are not included in an importance-based object ranking $R$, an input of \algoref{alg:drawing_sd}.

As shown in \tabref{tab:grouped_objects_count},
on average, \SI{50.2}{\percent} of the non-temporary objects are grouped by our technique (i.e., \SI{50.2}{\percent} of the objects are either template, hook, or delegate objects).

As shown in \tabref{tab:patterns_rate_in_code} and \tabref{tab:grouped_objects_count},
our technique uncovered many opportunities for object grouping.
In addition, our technique achieved high $F$ scores and \textit{Recall}s for all the subject systems, regardless of the differences in the rates of grouped objects (see \tabref{tab:performance_class_level}, \figref{fig:performance_ours_baseline_f1}, and \figref{fig:performance_ours_baseline_max_recall}).
In other words, objects corresponding to the important concepts tend to utilize template/hook structures or delegations.

\begin{table}[tb]
 \centering
 \caption{Numbers of grouped and non-grouped objects.
    The ``Grouped'' column shows the number of objects grouped by ``MP+D''.
    The ``Non-grouped'' column shows the number of non-grouped objects.
    The ``Rate of grouped'' column is the rate of the grouped objects to all the objects. 
    The number outside (resp.\ inside) the parentheses shows the number of non-temporary objects (resp.\ all the objects).
    }
 \label{tab:grouped_objects_count}
  \begin{tabular}{crrr}
   \hline
   Project & Grouped & Non-grouped & Rate of grouped [\%] \\
   \hline
   jpacman    & 147 \ \ \ (278)  & 409 \ \ \ (995) & 26.4 (21.8) \\
   JModeller  & 149 (4,625)      & 119 (2,368)     & 55.6 (66.1) \\
   wro4j      & 107 \ \ \ (240)  & 277 (1,264)     & 27.9 (16.0) \\
   JMeter     & 3,962 (4,543)    & 392 (1,019)     & 91.0 (81.7) \\
   \hline
   Average    & 1,091 (2,422)    & 299 (1,399)     & 50.2 (46.4) \\
   \hline
  \end{tabular}
\end{table}

\begin{framed}
 \noindent
 Numerous meta patterns and delegations are used in the subject systems.
  On average, \SI{33.2}{\percent} of all the method invocations were hook/delegate method invocations.
 \SI{50.2}{\percent} of all the non-temporary objects were grouped by our technique.
 There were many grouping opportunities found by our technique. Therefore, we expect our technique to have a wide range of application.
 Moreover, our technique achieved high quality object grouping, regardless of the rates of meta patterns and delegations used; objects of important concepts tended to utilize meta patterns or delegations.
\end{framed}

\revA{\subsubsection{Answer to $\textrm{RQ}_3$}}
\label{sec:experiment:answer_rq4}

\tabref{tab:runtime_overhead} shows the runtime overhead.
For each execution scenario,
 we measured the execution time five times, both with and without the logging codes for recording an execution trace.
We then calculated the average overhead.
We used an Intel Xeon E5-2620 v4 2.10GHz machine and assigned 16GB
of RAM to the heap of the Java VM.
We set the options for \textit{SELogger} as follows:
four background threads were used for writing trace data onto a disk;
the trace data was recorded in an uncompressed format.

\begin{table}[tb]
 \centering
 \caption{Runtime overhead.
 The ``Base'' (resp.\ ``With logging'') column shows the execution time without (resp.\ with) logging codes. The ``Overhead'' is calculated by $(\textrm{``With logging''} - \textrm{``Base''}) / \textrm{``Base''} \cdot 100$.
 }
 \label{tab:runtime_overhead}
  \begin{tabular}{crrrr}
    \hline
    Project & Base [\si{\second}] & With logging [\si{\second}] & Overhead [\si{\percent}] &\\
    \hline
    jpacman   & 9.84 & 13.76 & 39.9 \\
    JModeller & 5.15 & 10.24 & 98.9 \\
    wro4j     & 4.59 & 6.16  & 34.0 \\
    JMeter    & 4.97 & 22.06 & 344.1 \\
    \hline
    Average   & 6.14 &  16.54 & 129.2 \\
    \hline
  \end{tabular}
\end{table}

Our technique imposed a runtime overhead of \SI{129.2}{\percent} on average.
This overhead is relatively small compared with recent scalable
dynamic analysis techniques\tcite{Madsen:2016:icse:feedback_directed,Huang:2016:icse:scalable_thread_sharing},
 which incur runtime overhead of approximately \SIrange{100}{800}{\percent}.

Developers only need to execute an instrumented application once to produce the behavioral visualization.
Thus, in many cases, the overhead from our technique is expected
to prove acceptable in a development phase rather than in a production phase.

\begin{framed}
 \noindent
 Our technique imposed a runtime overhead of \SI{129.2}{\percent} on average, which is
 relatively small compared with recent dynamic scalable analysis
 techniques.
 In many cases, this overhead is expected to be acceptable in program comprehension tasks.
\end{framed}

$ $

\begin{figure*}[t]
 \centering
 \includegraphics[width=.77\textwidth]{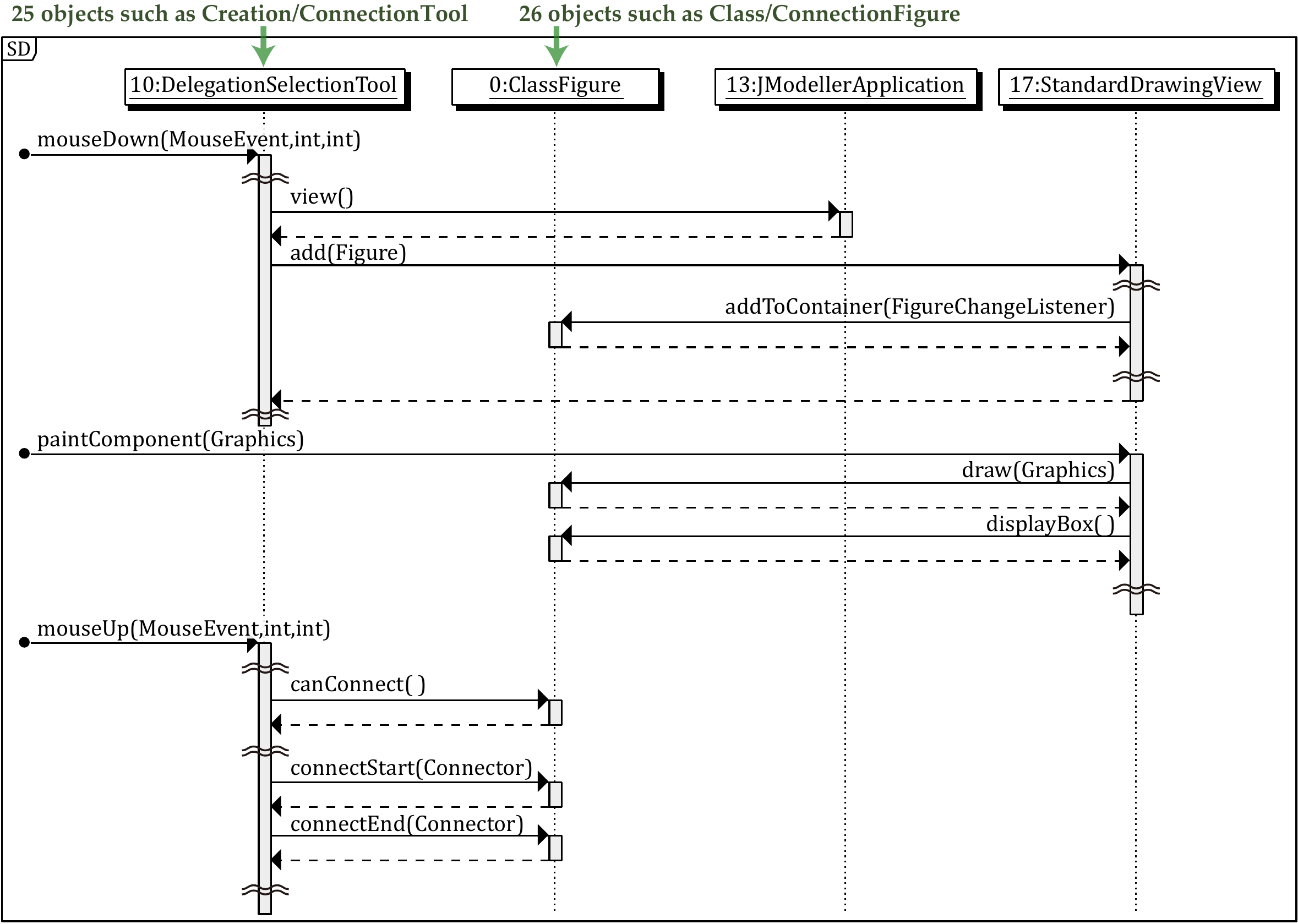}
 \caption{Portion of the resulting diagram from the \textit{JModeller} case.}
 \label{fig:result_sd_jmodeller}
\end{figure*}

Finally,
to facilitate reader understanding, 
we show an example covering a portion of the resulting (summarized) sequence diagram
from the \textit{JModeller} case in \figref{fig:result_sd_jmodeller}.
\figref{fig:result_sd_jmodeller} provides the knowledge for a behavioral overview of the subject system as follows.

When a user clicks on the canvas to add a new class figure,
 the tool objects (the lifeline named ``10:DelegationSelectionTool'') receive a mouse-down event (mouseDown(...)),
 and a message communicating the addition of a new figure (add(...)) is sent to the view objects (the lifeline named ``17:StandardDrawingView'').
Then, the view objects are set as containers of the newly added figure object (addToContainer(...)).

A paint-message (paintComponent(...)) is sent from library code.
Then, the view objects draw the figures in the view (draw(...)).

When a user clicks on the canvas to add a connector between two classes, 
 the tool objects receive a mouse-up event (mouseUp(...)).
 After testing whether it is possible to connect the two classes (canConnect(...)),
 the tool objects add a new connection (connectStart/End(...)).

In the resulting diagram, objects are abstracted at a concept level (for example, \textit{figure} (``0:ClassFigure'') or \textit{tool} (``10:DelegationSelectionTool'')), which is useful for comprehending the behavioral overview of the subject system.
The resulting diagram is expected to become a valuable tool for developers in an early stage of program comprehension.

\section{Threats to Validity}
\label{sec:threats_to_validity}

To improve the external validity, we used various types of applications in our experiment:
 a game application that periodically updates/renders the display according to the frame rate (jpacman);
 a modeling application written as an event-driven GUI program (JModeller); 
 a command line application that includes file processing (wro4j);
 a multi-threaded command line application that includes network accesses (JMeter).
However, because the number of subject systems is limited, 
it is unclear whether the results of our experiment can be generalized further.

\R{a-3-threats-1}{
Our algorithm groups objects involved in template-hook (and delegation) structures.
We cannot theoretically validate whether focusing only on those structures is sufficient for obtaining useful summarization.
Instead, we showed the following through our experiment.
\begin{itemize}
  \item Template-hooks and delegations are widely used in object-oriented systems: over 33\% of method invocations are hooks or delegations, and over 50\% of non-temporary objects are involved in hooks or delegations (\revA{\secref{sec:experiment:answer_rq2}}).
  This indicates our algorithm would have  \revA{many opportunities to summarize object behavior}.
  \item The resulting summarization has not deviated significantly from those described in execution scenarios and documents: F score and Recall are 0.670 and 0.793, resp (RQ$_1$).
  This means the level (degree) of our abstraction is neither too excessive nor too insufficient.
\end{itemize}
Focusing template-hook (and delegation) structures gives us numerous opportunities to group objects and a good abstraction level.
We therefore consider that our algorithm based only on template-hooks (and delegations) works effectively for obtaining behavioral overviews of object-oriented systems.
}

We defined the ground truths by ourselves.
To improve objectivity, we extracted the ground truths from the documents written by developers (except in the case of jpacman);
 however, the definitions of the ground truths might be incorrect.
 To mitigate the threat, we listed all the ground truths in this paper (\tabref{tab:ground_truths}) so that subsequent research can use and validate them.

We did not determine how much software maintenance task time is saved by our summarized sequence diagram;
 because this study focuses on the compactness of the diagram and the quality of the object grouping, an evaluation of time savings is outside the scope of this paper.
 Further studies are needed in order to evaluate the usefulness of summarized sequence diagrams in actual maintenance tasks.

\section{Related Work}
\label{sec:related_work}

\subsection{Coping with Large Execution Traces}
\label{sec:related_work:trace_summarization}

To improve program comprehension, testing, and formal verification,
 many studies have focused on recovering sequence diagrams\tcite{bennett08:_survey_and_evaluat_of_tool,Ghaleb:survey_sd:2018:JSEP,Lo:2011:ICECCS:succinctness_live_sd}.
The existing research has stated that because execution traces contain vast amounts of information,
 reverse-engineered sequence diagrams are often afflicted by scalability issues.

\subsubsection{Vertical Summarization of a Reverse-engineered Sequence Diagram}
\label{sec:related_work:vertical_summarization}

The primary approach for coping with the scalability issue
 is summarizing/abstracting the repetitive behavior in an execution trace\tcite{taniguchi05:_extrac_sequen_diagr_from_execut,myers10:_utiliz_debug_infor_to_compac,jayaraman:2016:SPE:compact_visualization}.
 Traces contain substantial amounts of repetitive behavior owing to iterative statements or recursive method calls in a program;
 thus, summarizing the repetitive behavior results in a significant reduction in the vertical size of a reverse-engineered sequence diagram.
To detect iterative behavior in a trace, for example, previous studies utilized debug information\tcite{myers10:_utiliz_debug_infor_to_compac} or 
proposed a regular expressions-based approach\tcite{jayaraman:2016:SPE:compact_visualization}.

An execution trace is often composed of several phases (tasks);
 thus, phase detection approaches\tcite{watanabe08:_featur_level_phase_detec_for,Pirzadeh:2011:ICECCS:gestalt_psychology_phase_detection,Pirzadeh:2011:ICSM:text_mining_phase_detection,Pirzadeh:2013:SCP:stratified_sampling_phase_detection}, which divide an entire trace into phases,
 are effective in reducing the vertical size of reverse-engineered sequence diagrams.
To detect phases, for example, previous studies examined the creation time of objects\tcite{watanabe08:_featur_level_phase_detec_for} or utilized text mining techniques\tcite{Pirzadeh:2011:ICSM:text_mining_phase_detection}.

Hamou-Lhadj and Lethbridge proposed a technique for removing unimportant methods (i.e., utilities) from a trace on the basis of the fan-in/out of each method\tcite{hamou-lhadj06:_summar_conten_of_large_traces}.
By removing unimportant methods (implementation details), they achieved a vertical and horizontal summarization of a reverse-engineered sequence diagram.

\subsubsection{Horizontal Summarization of a Reverse-engineered Sequence Diagram}
\label{sec:related_work:horizontal_summarization}

\R{b-3-category-1}{
Along with the vertical summarizations, 
horizontal summarizations of reverse-engineered sequence diagrams are an important factor in improving overall practicality.
Some existing trace summarization approaches perform horizontal summarization.
The technique proposed in this paper is also categorized as this type.
}

OGAN is a tool that visualizes interactions between objects related to two classes specified by a user\tcite{munakata09:_ogan}.
For each object, OGAN computes the dynamic interaction context, which is a set of classes that use/are-used-by the object.
Then, for each class, objects are clustered according to the equality of their dynamic interaction contexts.
OGAN selects an object from a cluster for each specified class, and visualizes interactions between those and related objects.
If developers already have a certain degree of knowledge about a subject system and know the class names of interest then OGAN is useful for obtaining a representative behavior of the classes.
Meanwhile, our technique is more suitable when developers do not know the class names that are important in comprehending the design overview of a subject system
 (e.g., in an early stage of program comprehension).

Dugerdil and Repond proposed a class clustering technique in which each cluster corresponds to a functional entity in a subject system\tcite{dugerdil10:_autom_gener_of_abstr_views}.
Their technique divides an entire trace into segments; then, for each loaded class, it creates a feature vector that represents the binary occurrence (presence (1) or absence (0)) of the class in each segment.
If the feature vectors of two classes are similar, those classes are gathered into the same cluster.

A trace, which is generated by exercising several features, consists of several functional phases.
Because the technique proposed by Dugerdil and Repond clusters classes from a functional perspective by segmenting a trace and calculating the similarities of the binary occurrence vectors,
 each of the constructed clusters tends to correspond to a functional phase.
As a consequence, the resulting summarized sequence diagram provides a highly abstracted behavioral view that depicts relationships (interactions/flows) among functional entities.
Their technique is therefore suitable for comprehending a long/complex execution scenario that exercises several features.
Compared with their technique, our technique aims at providing a finer-grained behavioral view.
As its input, our technique assumes a simpler trace that exercises a few features, and visualizes only important behavior (interactions among important objects).
Thus, our technique is suitable for comprehending how key objects behave in a simpler execution scenario that exercises a few features of interest.

\R{a-3-related-work-1}{
Toda et al.\ proposed an object-grouping technique based on the GoF design patterns\tcite{Toda:2013:APSEC_Companion:grouping_obj_design_pattern}.
The key idea of using design patterns for clustering is similar to our approach.
The technique by Toda et al.\ focused on the GoF design patterns, which are more specific than Pree's meta patterns.
Therefore, although the risk of grouping irrelevant objects into the same cluster is low, the number of opportunities for object grouping is quite small.
Our technique, which clusters objects based on more primitive design patterns,  \revA{finds more opportunities to summarize object behavior}.
}

\R{b-3-rel-1}{
\revB{
We also proposed similar approaches in our previous papers\tcite{Noda:COMPSAC2017:core_objects,noda:2018:IEICETrans:identifying_core}.
In the study\tcite{Noda:COMPSAC2017:core_objects}, we first identified core objects based on reference relations and access frequencies.
Then, grouping core-related objects based on lifetimes and reference relations, we obtained horizontally-summarized reverse-engineered sequence diagrams that depicted inter-group interactions among the core object groups.
The subsequent study\tcite{noda:2018:IEICETrans:identifying_core} presented a refined version of the core identification algorithm (described in \secref{sec:background:core_identification}) and produced better results.
}}

\R{b-3-rel-2}{
\revB{
We selected the study{\tcite{noda:2018:IEICETrans:identifying_core}} as the baseline in our experiment.
As described in {\secref{sec:rqs}}, the technique proposed in this paper leverages the core identification algorithm proposed in the baseline paper{\tcite{noda:2018:IEICETrans:identifying_core}}.
The difference between the proposed technique and the baseline lies in the object-grouping algorithms.
The proposed technique constructs important object groups based on Pree's meta patterns as described in {\secref{sec:proposed_technique:object_grouping}}.
Meanwhile},  \revB{for} each important (core) object $o_\textrm{imp}$, the baseline technique identifies a set of objects that are in a composition relation together with $o_\textrm{imp}$, and gathers those objects into the same group.
In \secref{sec:experiment}, we investigated the performance of the baseline technique, and
 showed that the technique proposed in this paper outperformed the baseline technique in terms of horizontal summarization of reverse-engineered sequence diagrams.
}

\subsubsection{Other Approaches}
\label{sec:related_work:other_approaches}

There are many other approaches for handling the massive amounts of information in a trace.

\R{b-3-category-2}{
Reticella extracted a partial behavioral view from an entire interaction log through \revB{a specialized dynamic slicing\tcite{noda:2012:IEICE:reticella}.
The slice (partial behavioral view) of a reverse-engineered diagram is calculated on a B-model (behavior model) event sequence.
As with traditional dynamic program slicing, the specialized slicing technique does not summarize (abstract) entire object interactions but pinpoints and extracts partial interactions of interest based on slicing criteria.}
Alimadadi et al.\ identified recurring structures (motifs) in a trace and represented the trace as a hierarchical view\tcite{Alimadadi:2018:ICSE:hierarchical_motif}.
Busany and Maoz proposed a technique that suggested a stopping criteria for a trace analysis (e.g., the sample size of a trace), such that the analysis results could be statistically trusted
 at a specified level of confidence\tcite{Busany:2016:ICSE:log_analysis_stats_guarantees}.
}

Some existing works proposed effective visualization tools: 
 a dedicated view for comprehending a large trace\tcite{cornelissen11:_contr_exper_for_progr_compr},
 an interactive tool that effectively explores a sequence diagram\tcite{sharp05:_inter_explor_of_uml_sequen_diagr},
 and a generic toolkit that provides a set of functionalities for smoothly exploring a massive-scale sequence diagram\tcite{Lyu:2018:ICPC:sdexplorer}.

\subsection{Detecting and Leveraging Design Patterns}
\label{sec:related_work:design_patterns_detection}

A number of design pattern detection techniques have been proposed\tcite{dong:2009:survey_design_pattern_detection,Ampatzoglou:2013:survey_gof_design_patterns}
While most existing techniques focus on GoF design patterns,
 some research detected and leveraged the Pree's meta patterns, which are the most primitive design patterns\tcite{hayashi08:_desig_patter_detec_by_using_meta_patter,posnett:2010:MSR:thex_meta_patterns}.
Because some GoF design patterns have similar structures, detecting these patterns produces some false positives.
On the other hand, because each Pree's meta pattern is structurally distinguishable, it is relatively easy to detect these patterns without false positives.

Detecting the meta patterns, Hayashi et al.\ reduced the computation and maintenance costs of the GoF design pattern detection logic\tcite{hayashi08:_desig_patter_detec_by_using_meta_patter}.
Posnett et al.\ proposed a meta pattern detector named Thex that worked on Java bytecode and is scalable to larger codebases \tcite{posnett:2010:MSR:thex_meta_patterns}.
Our technique also detects meta patterns for constructing object groups corresponding to concepts.

\section{Conclusion}
\label{sec:conclusion}

The behavior of an object-oriented program is visualizable as a reverse-engineered sequence diagram, which is a valuable tool for program comprehension;
 however, owing to the massive size of execution traces, reverse-engineered sequence diagrams often suffer from scalability issues.

To address the issues, in this paper we propose a trace summarization technique that reduces the horizontal size of reverse-engineered sequence diagrams.
Our technique constructs object groups in which each object corresponds to a concept in the domain of a subject system.
Then, given an importance-based object ranking, important object groups are identified.
By visualizing intergroup interactions only among the important groups, our technique generates a summarized version of a reverse-engineered diagram
 that depicts a behavioral overview of the subject system.

We evaluated our technique using traces generated from various types of open source software.
The results showed that our technique outperformed the state-of-the-art trace summarization technique in terms of horizontal summarization of reverse-engineered sequence diagrams.
Regarding the quality of object grouping, 
 our technique achieved an $F$ score (resp.\ a \textit{Recall}) of 0.670 (resp.\ 0.793) on average,
 under the condition of \#lifelines $< 30$ (which is acceptable for manual investigation). This is 0.249 (resp.\ 0.123) higher than that of the baseline technique.
Moreover, our technique imposed a runtime overhead of \SI{129.2}{\percent} on average, which is relatively small compared with recent scalable dynamic analysis techniques.
In many cases, the runtime overhead of our technique is expected to be acceptable in maintenance tasks.

Overall, 
our technique can recover a summarized sequence diagram that depicts a behavioral overview, while incurring a small runtime overhead.
The resulting summarized diagram is expected to be a valuable tool for developers in an early stage of program comprehension.

\begin{acknowledgment}
This work was partly supported by MEXT KAKENHI Grant Numbers JP15H02683, JP18H03221, and JP18J15087.
\end{acknowledgment}

\begin{biography}

\profile{Kunihiro NODA}{
received the B.Sc.\ degree in engineering and M.Sc.\ degree in information science from Nagoya University,
 and the Ph.D.\ degree in computer science from Tokyo Institute of Technology.
He had worked as a software engineer in an automotive company for around five years,
 and is currently working as a software engineering researcher in industry.
His research interests include program analysis, program comprehension, specification mining, automated program repair, and debugging.
}

\profile{Takashi KOBAYASHI}{
is an associate professor in the Department of Computer Science, School of
Computing, Tokyo Institute of Technology. His research interests
include software reuse, software design, software maintenance, program
analysis, software configuration management, Web-services compositions,
workflow, multimedia information retrieval, and data mining. He received
B.Eng., M.Eng.\, and Dr.Eng.\ degrees in computer science from Tokyo
Institute of Technology in 1997, 1999, and 2004, respectively. He is a
member of the JSSST, IPSJ, DBSJ, IEEE-CS, and ACM.
}

\profile{Kiyoshi AGUSA}{
received the B.Sc.\ and M.Sc.\ degrees in electrical engineering, and Ph.D.\ degree in computer science from Kyoto University in 1970, 1972 and 1980, respectively.
He is a president, vice-chairman of Advanced Science, Technology \& Management Research Institute of KYOTO (ASTEM).
His research interests lie in the field of software engineering, especially programming environments, and requirements.
}

\end{biography}

\end{document}